\documentclass[12pt]{article}
\usepackage{bm}
\usepackage{ulem}
\usepackage{amsmath,amssymb}
\usepackage{graphicx,psfrag,epsf}
\usepackage{enumerate}
\usepackage{natbib}
\usepackage{multirow} 
\usepackage{mathtools}
\usepackage{url} 

\newtheorem{theorem}{Theorem}

\newtheorem{lemma}{Lemma}

\newtheorem{corollary}{Corollary}
\newtheorem{assumption}{Assumption}

\newtheorem{example}{Example}

\newtheorem{remark}{Remark}

\newcommand{\blind}{0}

\def\argmin{\mathop{\rm argmin}}

\def\max{\mathop{\rm max}}

\usepackage{color}
\newcommand{\mht}[1]{{\color{blue}#1}}

\usepackage{color}
\newcommand{\szl}[1]{{\color{green}#1}}

\begin{document}

\def\spacingset#1{\renewcommand{\baselinestretch}%
{#1}\small\normalsize} \spacingset{1}


\if0\blind
{
  \title{\bf Sparsity learning via structured functional factor augmentation}
  \author{\small Hanteng Ma\\
    \small School of Statistics and Data Science, Shanghai University of Finance and Economics\\
    \small Ziliang Shen \\
    \small School of Statistics and Data Science, Shanghai University of Finance and Economics\\
    \small Xingdong Feng \\
    \small School of Statistics and Data Science, Shanghai University of Finance and Economics\\
    \small Xin Liu\thanks{
    The authors gratefully acknowledge \textit{please remember to list all relevant funding sources in the unblinded version}}\hspace{.2cm} \\
    \small School of Statistics and Data Science, Shanghai University of Finance and Economics}
  \maketitle
} \fi

\if1\blind
{
  \bigskip
  \bigskip
  \begin{center}
    {\LARGE\bf 
   Sparsity learning via structured functional factor augmentation}
\end{center}
  \medskip
} \fi

\bigskip
\begin{abstract}
As one of the most powerful tools for examining the association between functional covariates and a response, the functional regression model has been widely adopted in various interdisciplinary studies. Usually, a limited number of functional covariates are assumed in a functional linear regression model. Nevertheless, correlations may exist between functional covariates in high-dimensional functional linear regression models, which brings significant statistical challenges to statistical inference and functional variable selection. In this article, a novel functional factor augmentation structure (fFAS) is proposed for multivariate functional series, and a multivariate functional factor augmentation selection model (fFASM) is further proposed to deal with issues arising from variable selection of correlated functional covariates. Theoretical justifications for the proposed fFAS are provided, and statistical inference results of the proposed fFASM are established. Numerical investigations support the superb performance of the novel fFASM model in terms of estimation accuracy and selection consistency.

\end{abstract}

\noindent%
\vfill

\newpage
\spacingset{1.45} 
\section{Introduction}
\label{sec:intro}
Functional data, usually referred to as a sequential collection of instances over time with serial dependence, have been widely observed in various scenarios from different scientific disciplines, such as earth, medical, and social sciences \citep{peng2005seasonal,centofanti2021functional}. Instead of employing conventional time series modeling techniques, an underlying trajectory is usually assumed for a functional process and sampled at a set of time points in some interval, so that statistical modeling and inference are feasible, including estimation of the mean process and its intrinsic covariance structure, trajectory recovery, and prediction of a functional; see, for example, \cite{ramsay2005functional, yao2005functional, ferraty2006nonparametric, hall2007methodology, hormann2010weakly, cuevas2014partial, hormann2015dynamic, aneiros2022functional, petersen2024mean}, among others. 

Conventionally, functional data analysis has focused on a single or fixed number of separate functional series, while recent studies have explored associations between multiple functional series, and the most widely adopted model may be the functional linear model, which assumes to (linearly) link a response with a bunch of covariates and at least one functional covariate by using unknown functional coefficient curves. Such a model offers a statistical tool to infer linear association between the response and functional covariates. Several studies have attempted along this path, employing parametric, non-parametric, and semi-parametric models; see, for instance,  \cite{cardot2003spline, yao2005functional, li2007rates, muller2008functional, yuan2010reproducing, chen2011single, kong2016partially, chen2022functional}, among others. 

However, two main issues arise in multivariate functional data analysis in practice. One is that functional series may be correlated with each other, especially in high dimensions or less dense observation cases, while the current functional linear model analysis may typically assume independence between multiple functional covariates. To address this problem, some efforts are paid, for example, to assume pairwise covariance structures for functional covariates explicitly \citep{chiou2016pairwise}. However, these methods turn out to be less robust or accurate due to lack of flexibility and computational efficiency.  Alternatively, one can describe such associations by using a factor model with lower ranks \citep{bai2003inferential, fan2020factor}, where all functional covariates are assumed to share a finite number of certain latent functional factors \citep{castellanos2015multivariate}, without assuming any explicit correlation structure for multivariate functional covariates. A few studies have touched the so-called functional factor models, e.g., \cite{hays2012functional, chen2021dynamic, yang2024robust}, while these methods provide limited exploration of common functional factors shared by functional covariates. To the best of our knowledge, very limited research has focused on how to capture common associations for multivariate functional series efficiently and effectively with inferential justifications from a statistical perspective. 

Another issue is how to select useful functional covariates in multivariate functional linear regression models when correlations between functional covariates exist, which is also frequently needed in practice, such as Internet of Things (IoT) data analysis \citep{gonzalez2019methodology}, antibiotic resistance prediction \citep{JIMENEZ2020101818}, nonsmall cell lung cancer (NSCLC) clinical study \citep{fang2015feature}, and stock market forecasting \citep{nti2019random,htun2023survey}. A common way to achieve functional variable selection is to impose a penalty on the corresponding functional coefficient curves in a group-wise manner, using popular penalties such as the lasso \citep{tibshirani1996regression}, SCAD \citep{fan2001variable} and MCP \citep{zhang2010nearly}. A few studies have touched on functional variable selection in certain scenarios; see, for example, \cite{MATSUI20113304, kong2016partially,aneiros2022functional}, among others. However, when correlation exists between functional covariates, such a strategy turns out to be less accurate in functional coefficient estimation and fails to capture the truly useful ones, as demonstrated in simulation studies in this article. Consequently, it remains statistically challenging how to select useful functional covariates with selection consistency in multivariate functional linear models with correlated functional covariates in high dimensions.

Inspired by the challenges above, we firstly propose a novel functional factor augmentation structure (fFAS) to capture associations for correlated multivariate functional data, and further propose a functional factor augmentation selection model (fFASM) to achieve selection of functional covariates in high dimensions when correlations exist between functional covariates using the penalized method. Not only is the correlation addressed without assuming an explicit covariance structure, but theoretical properties of the estimated functional factors are also established. Further, pertaining to the correlated functional covariates, the proposed fFASM method successfully captures the useful functional covariates simultaneously in the context of functional factor models. Numerical studies on both simulated and real datasets support the superior performance of the proposed fFASM. The main contributions of the proposed method are two-fold as follows.
\begin{itemize}
    \item We propose a feasible time-independent functional factor augmentation structure (fFAS) for functional data, and establish theoretical justifications for statistical inference, revealing valuable insights into current literature. Also, the impact from truncation inherent in functional data analysis to the proposed fFAS is discussed, and solutions are provided to enhance robustness and applicability of our model. A key result is how the difference may be addressed and controlled by comparing the true factor model and the estimated one when using truncated expansion.
    \item A multivariate functional linear regression model with correlated functional covariates is proposed, by employing the proposed functional structural factor. As correlations often exist in multiple functional covariates in high dimensions but may be difficult to estimate, the proposed fFASM method decomposes the functional covariates into two parts with low correlations with each other, and simultaneously expands the dimensions of parameters to be estimated. By this way our approach improves the performance of functional variable selection.
\end{itemize}

The rest of the paper is organized as follows. Section \ref{sec:fFs} introduces a novel fFAS for functional processes with its statistical properties, and the fFASM is further proposed in Section \ref{sec:fFASM} for correlated functional covariates and functional variable selection in detail with theoretical justifications. Section \ref{sec:Simulation} employs simulated data to examine the proposed method in various scenarios, and Section \ref{sec:Real} presents its applications on two sets of real-world datasets. Section \ref{sec:Conclusion} concludes the article with discussions.

\textbf{Notation}: 
$\bm{I}_n$ denotes the $n \times n$ identity matrix;
$\bm{0}$ refers to the $n \times m$ zero matrix;
$\bm{0}_n$ and $\bm{1}_n$ represent the all-zero and all-one vectors in $\mathbb{R}^n$, respectively. For a vector $\bm{a}$, $\|\bm{a}\|_2$ denotes its Euclidean norm. For a vector $\bm{v} \in \mathbb{R}^p$ and $S \subseteq[p]$, denote $\bm{v}_S=\left(\bm{v}_i\right)_{i \in S}$ as its sub-vector. For a matrix $\bm{M} \in \mathbb{R}^{n \times m}, I \subseteq[n]$ and $J \subseteq[m]$, define $\bm{M}_{I J}=\left(\bm{M}_{i j}\right)_{i \in I, j \in J}, \bm{M}_{I .}=\left(\bm{M}_{i j}\right)_{i \in I, j \in[m]}$ and $\bm{M}_{. J}=\left(\bm{M}_{i j}\right)_{i \in[n], j]}$, and its matrix entry-wise max norm is denoted as $\|\bm{M}\|_{\max }=\max _{i, j}\left|M_{i j}\right|$, and $\|\bm{M}\|_F$ and $\|\bm{M}\|_p$ as its Frobenius and induced $p$-norms, respectively. Denote $\lambda_{\min }(\bm{M})$ as the minimum eigenvalue of $\bm{M}$ if it is symmetric. Let $\nabla$ and $\nabla^2$ be the gradient and Hessian operators.
For $f: \mathbb{R}^p \rightarrow \mathbb{R}$ and $I, J \subseteq[p]$, define $\nabla_I f(x)=(\nabla f(x))_I$ and $\nabla_{I J}^2 f(x)=\left(\nabla^2 f(x)\right)_{I J}$. $N(\boldsymbol{\mu}, \boldsymbol{\Sigma})$ refers to the normal distribution with mean $\boldsymbol{\mu}$ and covariance matrix $\boldsymbol{\Sigma}$.

\section{A novel functional factor augmentation structure}\label{sec:fFs}
\subsection{Functional observations and its expansion}
To start with, suppose that $G$ functional processes are observed and collected sequentially in a sample $\{W_{i}^{(g)}(t_{ij}^{(g)}), t_{ij}^{(g)}\in\mathcal{T};~g=1,\ldots, G,~i=1,\ldots,n,~j=1,\ldots,n_{i}^{(g)}\}$ for the $g$-th functional process from the $i$-th subject at time $t_{ij}$ in a certain time interval $\mathcal{T}$, and we use $W_i^{(g)}(t_j)$ for abbreviation without confusion. Usually in the analysis of a single functional process, $W_i^{(g)}(t_j)$ is assumed to consist of the underlying process $X_i^{(g)}(t_j)$ and an independent noise $\varepsilon_i^{(g)}(t_j)$, i.e., $W_i^{(g)}(t_j) = X_i^{(g)}(t_j)+\varepsilon_i^{(g)}(t_j)$, where {$X_i^{(g)}(\cdot)$ and $\varepsilon_i^{(g)}(\cdot)$ are assumed to be identically and independently distributed (i.i.d.) with a mean function and a covariance structure, respectively, and $\mathbb{E}(\varepsilon_i^{(g)}(t)) = 0,\mathbb{E}(X_i^{(g)}(t)) = \mu(t)$.} {For convenience, we centralize these functional processes and still use the notation $X_i^{(g)}(\cdot)$ to represent them.} To recover the functional trajectory $X_i^{(g)}(t)$, it is usually smoothed by assuming an expansion over a set of pre-specified orthogonal basis functions $\{\phi_{0j}^{(g)}(\cdot),j = 1,\ldots,m_g \}$ as 
\begin{equation}\label{eq:KLexpansion}
X_i^{(g)}(t) = \sum_{j=1}^{m_g} a_{0,ij}^{(g)}\phi_{0j}^{(g)}(t) + e_{0i}^{(g)}(t) = \bm{a}_{0,i}^{(g)}{}^\top \bm{\phi}_{0 }^{(g)}(t) + e_{0i}^{(g)}(t), 
\end{equation}
where $\bm{a}_{0,i}^{(g)} = ( a_{0,i1}^{(g)},\ldots,a_{0,im_g}^{(g)} )^\top$ are the time-invariant coefficients, $\bm{\phi}_{0 }^{(g)}(t) = (\phi_{01}^{(g)}(t),\ldots,\phi_{0m_g}^{(g)}(t))^\top$, and $e_{0i}^{(g)}(\cdot)$ is the residual orthogonal to $\phi_{0j}^{(g)}(\cdot)$. As in practice, $\bm{\phi}_{0 }^{(g)}(\cdot)$ is usually unknown, an identifiable estimator $\widehat{\bm{a}}_{0,i}^{(g)}$ can be obtained under some conditions, shown in Lemma \ref{lemma:basis}.

\begin{lemma} \label{lemma:basis}For $X_i^{(g)}(\cdot)$ which has the following properties,
\begin{itemize}
    \item $\mathbb{E}(X_i^{(g)}(t)) = \mathbb{E}(a_{0,ij}^{(g)}) = \mathbb{E}(a_{0,ij}^{(g)}a_{0,ik}^{(g)}) = \mathbb{E}(e_{0,i}^{(g)}(t)) = \mathbb{E}(a_{0,ij}^{(g)}e_{0i}^{(g)}(t)) = 0$, where $k \neq j$,
    \item $\mathbb{E}(X_i^{(g_1)}(s)e_{0i}^{(g_2)}(t)) = 0$ for $g_1 \neq g_2$, $g_1,g_2 = 1,\ldots,G$, $s,t\in\mathcal{T}$,
    \item $\operatorname{Var}(a_{0,i1}^{(g)}) > \operatorname{Var}(a_{0,i2}^{(g)}) > \ldots > \operatorname{Var}(a_{0,im_g}^{(g)})$,
    \item all the eigenvalues of $\operatorname{Cov}(e_{0,i}^{(g)}(s),e_{0,i}^{(g)}(t))$ are less than {$\lambda_{\min}(\operatorname{Cov}(\bm{a}_{0,i}^{(g)}))$}, where $\lambda_{\min}(\operatorname{Cov}(\bm{a}_{0,i}^{(g)}))$ is the smallest eigenvalue of $\operatorname{Cov}(\bm{a}_{0,i}^{(g)})$,
\end{itemize}
      where the $j$-th eigenfunction $\gamma_j^{(g)}(\cdot)$ of $\operatorname{Cov}(X_i^{(g)}(s),X_i^{(g)}(t))$ is $\phi_{0j}^{(g)}(\cdot)$, so that the $j$-th functional score $a_{ij}^{(g)} \coloneqq \int_{\mathcal{T}} X_i^{(g)}(t) \gamma_j^{(g)}(t) dt = a_{0,ij}^{(g)}$.
\end{lemma}

{
\begin{remark}
 Lemma \ref{lemma:basis} states $\bm{a}_{i}^{(g)}=({a}_{i1}^{(g)},\ldots, {a}_{i{m_g}}^{(g)})^{\top}$ is a reasonable approximation of $\bm{a}_{0,i}^{(g)}$. All eigenvalues of $\operatorname{Cov}(e_{0,i}^{(g)}(s),e_{0,i}^{(g)}(t))$ are required to be less than $\lambda_{\min}(\operatorname{Cov}(\bm{a}_{0,i}^{(g)}))$, indicating that segregation can be conducted based on contribution to the variance of $X_i^{(g)}(t)$.
\end{remark}
}

The whole recovery process can be achieved by the popular functional principal component analysis (fPCA) with the Karhunen-Lo\'eve (KL)  expansion \citep{yao2005functional}.

\subsection{A functional factor augmentation structure}
As correlations may exist between multivariate functional processes, we propose a functional factor augmentation structure (fFAS) to address the issue. Consider a simplified scenario with only two correlated functional processes, $X_i^{(1)}(t)$ and $X_i^{(2)}(t)$, generated by
\begin{align}\label{eq:expansion}
    X_i^{(1)}(t) &=  \bm{a}_{0,i}^{(1)}{}^\top\cdot\bm{\phi}_{0}^{(1)}(t) + e_{0i}^{(1)}(t),\\
    X_i^{(2)}(t) &= \bm{a}_{0,i}^{(2)}{}^\top\cdot\bm{\phi}_{0}^{(2)}(t) + e_{0i}^{(2)}(t), \notag
\end{align}
where $e_{0i}^{(1)}(t)$ and $e_{0i}^{(2)}(t)$ are independent of each other. Assume $\bm{a}_{0,i}^{(g_1)}$ and $e_{0i}^{(g_2)}(t)$ are uncorrelated for $g_1,g_2 = 1,\ldots,G$, where $G=2$, and hence the correlation between $X_i^{(1)}(t)$ and $X_i^{(2)}(t)$ arises only from that between $\bm{a}_{0,i}^{(1)}$ and $\bm{a}_{0,i}^{(2)}$. To capture such a correlation, it is assumed that each $\bm{a}_{0,i}^{(g)}$ shares $K$ common underlying factors using a linear combination
\begin{align}\label{eq:factor}
    \left(
    \begin{array}{ll}
         \bm{a}_{0,i}^{(1)}  \\
         \bm{a}_{0,i}^{(2)} 
    \end{array}
    \right) = 
    \left(
    \begin{array}{ll}
         B^{(1)}  \\
         B^{(2)} 
    \end{array}
    \right)\bm{f}_i + 
    \left(
    \begin{array}{ll}
         \bm{u}_i^{(1)}  \\
         \bm{u}_i^{(2)} 
    \end{array}
    \right) = \bm{B} \bm{f}_i + \bm{u}_i,
\end{align}
where $\bm{B} = \left( B^{(1)}{}^\top ,B^{(2)}{}^\top\right)^\top$ is a $p\times K$ factor loading matrix with $p = \sum_{g=1}^{G}m_g$, $\bm{f}_i$ is a $K\times 1$ vectorized latent factors, and $\bm{u}_i$ is an idiosyncratic component independent of $\bm{f}_i$ which carries a weak correlation. Note that the covariance $C_{\bm{a}_0} := \operatorname{Cov}(\bm{a}_{0,i}) = \operatorname{Cov}(\bm B\bm{f}_i + \bm{u}_i) = \bm{B}\cdot\operatorname{Cov}(\bm{f}_i)\cdot\bm{B}^\top + \operatorname{Cov}(\bm{u}_i)$. For model identifiability, it is assumed that $\operatorname{Cov}(\bm{f}_i) = \bm{I}_K$ and $\operatorname{Cov}(\bm{u}_i) = \omega\Lambda_u$ with a $p\times p$ matrix $\Lambda_u$, where $\Vert\Lambda_u\Vert_{\max} \leq C_u$, a constant depending on the distribution of $\bm{u}_i$. By the spectral decomposition, $\bm B=\left(\sqrt{\lambda_1^{(B)}} \bm\xi_1^{(B)}, \ldots, \sqrt{\lambda_K^{(B)}}\bm\xi_K^{(B)}\right)$ with the eigenvalues $\lambda_1^{(B)} > \ldots > \lambda_K^{(B)}$ of $\bm{BB}^\top$ and the corresponding eigenvectors $\{\bm\xi_k^{(B)}, k = 1,\ldots,K\}$. In this way, $X_i^{(1)}(t)$ and $X_i^{(2)}(t)$ in \eqref{eq:expansion} can be expanded as
\begin{align}\label{eq:basis3}
    X_i^{(1)}(t) &= {\bm{\phi}_0^{(1)}(t)}{}^{\top} (B^{(1)}\bm{f}_i + \bm{u}_i^{(1)}) + e_{0i}^{(1)}(t) = {{\tilde{\bm{\phi}_0}}{}^{(1)}(t)}{}^{\top}\bm{f}_i + {\bm{\phi}_0^{(1)}(t)}{}^{\top}\bm{u}_i^{(1)} + e_{0i}^{(1)}(t),\\
    X_i^{(2)}(t) &= {\bm{\phi}_0^{(2)}(t)}{}^{\top} (B^{(2)}\bm{f}_i + \bm{u}_i^{(2)}) + e_{0i}^{(2)}(t) = {{\tilde{\bm{\phi}_0}}{}^{(2)}(t)}{}^{\top}\bm{f}_i + {\bm{\phi}_0^{(2)}(t)}{}^{\top}\bm{u}_i^{(2)} + e_{0i}^{(2)}(t),\notag
\end{align}
where ${{\tilde{\bm{\phi}_0}}{}^{(g)}(t)}{}^{\top}={\bm{\phi}_0^{(g)}(t)}{}^{\top} B^{(g)}$. This indicates $X_i^{(g)}(t)$ can be decomposed into two correlated parts, namely, the functional factor part $\boldsymbol{f}_i$ and the weakly correlated part ${\bm{\phi}_0^{(g)}(t)}{}^{\top}\bm{u}_i^{(g)}$, plus an independent error term. We provide an example to illustrate such a fFAS when two functional covariates share a linear structure.
\begin{example}
    Suppose $X_i^{(1)}{ (\cdot)}$ and $X_i^{(2)}{ (\cdot)}$ are associated with a linear structure as
\begin{align*}
    X_i^{(2)}{(t)} = \mathbb{E}(X_i^{(2)}(t)|X_i^{(1)}{(t)}) +\epsilon_i(t) = \int_{\mathcal{T}} \beta(s,t)X_i^{(1)} (s)ds+\epsilon_i{(t)},
\end{align*}
with $\beta(s,t) = \sum_{k=1}^{\infty}\sum_{m=1}^{\infty}\frac{\mathbb{E}(a_{0,im}^{(1)}a_{0,ik}^{(2)})}{\mathbb{E}(a_{0,im}^{(1)}{}^2)}\phi_{0k}^{(2)}(t)\phi_{0m}^{(1)}(s)$. Then it is obtained that
\begin{align*}
    \int_{\mathcal{T}} \beta(s,t)X_i^{(1)}(s)ds &= \int_{\mathcal{T}} \beta(s,t) \sum_{j=1}^{\infty} a_{0,ij}^{(1)}\phi_{0j}^{(1)}(s)ds \\
    &= \int_{\mathcal{T}}\sum_{k=1}^{\infty}\sum_{m=1}^{\infty}\frac{\mathbb{E}(a_{0,im}^{(1)}a_{0,ik}^{(2)})}{\mathbb{E}(a_{0,im}^{(1)}{}^2)}\phi_{0k}^{(2)}(t)\phi_{0m}^{(1)}(s) \sum_{j=1}^{\infty} a_{0,ij}^{(1)}\phi_{0j}^{(1)}(s)ds\\
    &= \sum_{k=1}^{\infty}\left(  \sum_{j=1}^{\infty} \frac{\mathbb{E}(a_{0,ij}^{(1)}a_{0,ik}^{(2)})}{\mathbb{E}(a_{0,ij}^{(1)}{}^2)}a_{0,ij}^{(1)}\right)\phi_{0k}^{(2)}(t),
\end{align*}
so that $a_{0,ik}^{(2)} \approx \sum_{j=1}^{\infty} \frac{\mathbb{E}(a_{0,ij}^{(1)}a_{0,ik}^{(2)})}{\mathbb{E}(a_{0,ij}^{(1)}{}^2)}a_{0,ij}^{(1)} = \sum_{j=1}^{m_1} \frac{\mathbb{E}(a_{0,ij}^{(1)}a_{0,ik}^{(2)})}{\mathbb{E}(a_{0,ij}^{(1)}{}^2)}a_{0,ij}^{(1)} + \sum_{j=m_1+1}^{\infty} \frac{\mathbb{E}(a_{0,ij}^{(1)}a_{0,ik}^{(2)})}{\mathbb{E}(a_{0,ij}^{(1)}{}^2)}a_{0,ij}^{(1)}$. With a truncation of the first $m_1$ and $m_2$ scores in $X_i^{(1)}(t)$ and $X_i^{(2)}(t)$, and combining them as ${\bm{a}_{0,i}}=({\bm{a}_{0,i}^{(1)}}^{\top},{\bm{a}_{0,i}^{(2)}}^{\top})^{\top}$, and a matrix $\bm{E}$ with elements $E_{kj} = \frac{\mathbb{E}(a_{0,ij}^{(1)}a_{0,ik}^{(2)})}{\mathbb{E}(a_{0,ij}^{(1)}{}^2)}$ and a diagonal matrix $\Lambda$ with elements $\Lambda_{jj} = \mathbb{E}(a_{0,ij}^{(1)}{}^2)$, it is easily obtained that 
\begin{equation*}
    \left(
    \begin{array}{ll}
         \bm{a}_{0,i}^{(1)}  \\
         \bm{a}_{0,i}^{(2)} 
    \end{array}
    \right) \approx 
    \left(
    \begin{array}{cc}
         \bm{I}_{m_1}  \\
         \bm{E} 
    \end{array}
    \right) \bm{a}_{0,i}^{(1)} =
    \left(\left(
    \begin{array}{cc}
        \bm{I}_{m_1}  \\
        \bm{E} 
    \end{array}
    \right) \Lambda^{\frac{1}{2}}\right) \left(\Lambda^{-\frac{1}{2}} \bm{a}_{0,i}^{(1)}\right)= \left(\left(
    \begin{array}{cc}
         \bm{I}_{m_1}  \\
         \bm{E}  
    \end{array}
    \right) \Lambda^{\frac{1}{2}} \bm{P}\right) (\bm{P}^{\top}\Lambda^{-\frac{1}{2}} \bm{a}_{0,i}^{(1)}),
\end{equation*}
where $\bm{P}$ is an orthogonal matrix, so that $\left(\left(
    \begin{array}{cc}
         \bm{I}_{m_1}  \\
         \bm{E}  
    \end{array}
    \right) \Lambda^{\frac{1}{2}} \bm{P}\right)^{\top}\left(\left(
    \begin{array}{cc}
         \bm{I}_{m_1}  \\
         \bm{E} 
    \end{array}
    \right) \Lambda^{\frac{1}{2}} \bm{P}\right)$ is diagonal, and $\bm{f}_i = \bm{P}^{\top}\Lambda^{-\frac{1}{2}} \bm{a}_{0,i}^{(1)}$ with $\operatorname{Cov}(\bm{f}_i) = \bm{I}_{m_1}$. Note that no matter what values of $m_1$ and $m_2$ are in practice, one can always obtain such a functional factor augmentation structure. 
    
\if{
In fact, we add nuisance term $U_i$
\begin{align*}
    X_i^{(2)}(t) = \int_{\mathcal{T}} \beta(s,t)X_i^{(1)}(s)ds + U_i(t),
\end{align*}
with $U_i(t) = \sum_ju_{ij}\eta_j(t)$ is independent of $X_i^{(1)}(t)$, then we can expand $X_i^{(2)}(t)=\sum_{j=1}^m a_{ij}^{(2)}\phi_j^{(2)} + e_i^{(2)} + \sum_ju_{ij}\eta_j(t)$, and the most of information in $X_i^{(2)}(t)$ is preserved by factor $\bm{f}_i = P^{\top}\Lambda^{-\frac{1}{2}} \bm{a}_i^{(1)}$.}
\fi
\end{example}

Next, we consider a more general case where the structure $\bm{a}_{0,i}^{(g)}$ contains correlations for $g = 1, \ldots, G$, where $\mathbb{E}(a_{0,ij}^{(g)}a_{0,ik}^{(g)}) \neq 0 
~(\text{for all }0<j<k<m_{g})$. Then, Lemma \ref{lemma:basis2} shows the relation between $\bm{a}_{i}^{(g)}$ and $\bm{a}_{0,i}^{(g)}$ in this circumstance that $\bm{a}_{i}^{(g)}$ is still an approximation of $\bm{a}_{0,i}^{(g)}$ by imposing an orthogonal rotation induced by the basis functions in K-L expansion.
\begin{lemma} \label{lemma:basis2}
    Under the conditions in Lemma \ref{lemma:basis} without $\mathbb{E}(a_{0,ij}^{(g)}a_{0,ik}^{(g)}) = 0$, there exists an orthogonal matrix $P^{(g)}$, such that $ \bm{\phi}_{0}^{(g)}(\cdot) = P^{(g)} ({\gamma}_{1}^{(g)}(\cdot),\ldots,{\gamma}_{m_g}^{(g)}(\cdot))^\top$, and
    \begin{align*}
    \left(
        \begin{array}{ll}
         a_{0,i1}^{(g)}  \\
         a_{0,i2}^{(g)} \\
         \vdots \\
         a_{0,im_g}^{(g)} 
    \end{array}
    \right)
    = P^{(g)} \left(
        \begin{array}{ll}
         a_{i1}^{(g)}  \\
         a_{i2}^{(g)} \\
         \vdots \\
         a_{im_g}^{(g)} 
    \end{array}
    \right).
    \end{align*} 
Furthermore, denote $\bm{a}_i^\top = (\bm{a}_i^{(1)}{}^\top,\ldots,\bm{a}_i^{(G)}{}^\top)$ and $\bm{a}_{0,i}^\top = (\bm{a}_{0,i}^{(1)}{}^\top,\ldots,\bm{a}_{0,i}^{(G)}{}^\top)$. Then 
\begin{align*}
 \operatorname{Cov}(\bm{a}_{i}) = 
    \begin{pmatrix}
 P^{(1)} & \bm{0} & \cdots  & \bm{0}\\
  \bm{0}&  P^{(2)}& \cdots & \bm{0}\\
  \vdots& \vdots &  \ddots& \vdots\\
  \bm{0}& \bm{0} & \cdots &P^{(G)}
\end{pmatrix}^\top \operatorname{Cov}(\bm{a}_{0,i}) \begin{pmatrix}
 P^{(1)} & \bm{0} & \cdots  & \bm{0}\\
  \bm{0}&  P^{(2)}& \cdots & \bm{0}\\
  \vdots& \vdots &  \ddots& \vdots\\
  \bm{0}& \bm{0} & \cdots &P^{(G)}
\end{pmatrix},
\end{align*} 
where
\begin{align*}
    {\operatorname{Cov}(\bm{a}_{i})} = \begin{pmatrix}
 \Sigma_{m_1} & \Sigma_{m_1m_2} & \cdots  & \Sigma_{m_1m_G}\\
  \Sigma_{m_2m_1}&  \Sigma_{m_2}& \cdots & \Sigma_{m_2m_G}\\
  \vdots& \vdots &  \ddots& \vdots\\
  \Sigma_{m_Gm_1}& \Sigma_{m_Gm_2} & \cdots &\Sigma_{m_G}
\end{pmatrix},
\end{align*}
and 
\begin{align*}
    {\operatorname{Cov}(\bm{a}_{0,i})} = \begin{pmatrix}
 \Sigma_{0,m_1} & \Sigma_{0,m_1m_2} & \cdots  & \Sigma_{0,m_1m_G}\\
  \Sigma_{0,m_2m_1}&  \Sigma_{0,m_2}& \cdots & \Sigma_{0,m_2m_G}\\
  \vdots& \vdots &  \ddots& \vdots\\
  \Sigma_{0,m_Gm_1}& \Sigma_{0,m_Gm_2} & \cdots &\Sigma_{0,m_G}
\end{pmatrix}.
\end{align*}

\end{lemma}

\begin{remark}
Lemma \ref{lemma:basis2} indicates that if there is a fFAS on $\bm{a}_{0,i}$, then there will be a fFAS on $\bm{a}_{i}$ by the fact that
\begin{align}\label{eq:ai}
    \bm{a}_{i} = \begin{pmatrix}
 P^{(1)} & \bm{0} & \cdots  & \bm{0}\\
  \bm{0}&  P^{(2)}& \cdots & \bm{0}\\
  \vdots& \vdots &  \ddots& \vdots\\
  \bm{0}& \bm{0} & \cdots &P^{(G)}
\end{pmatrix}^\top \bm{a}_{0,i} = \bm{P}^\top \bm{a}_{0,i} = \bm{P}^\top\bm{B} \bm{f}_i + \bm{P}^\top\bm{u}_i,
\end{align}
and treating $\bm{P}^\top\bm{B}$ and $\bm{P}^\top\bm{u}_i$ as the updated loading matrix and the idiosyncratic component in \eqref{eq:factor}, respectively, where $\bm{P}$ is still an orthogonal matrix. More specifically, if $\bm{a}_{0,i}$ has a fFAS induced by $\bm{f}_i$, $\bm{a}_{i}$ also has such a fFAS with the same factor $\bm{f}_i$. Consequently, even if the functional covariates are correlated with each other, we can still use the functional scores $\bm{a}_{i}$ to estimate $\bm{a}_{0,i}$ as if they were uncorrelated. 
\end{remark}
To further get the estimates of the loading matrix ${\bm{B}}$ and the functional factors $\bm{F} = \left(\bm{f}_1 ,\ldots ,\bm{f}_n \right)^\top \in \mathbb{R}^{n \times K}$ after obtaining the score matrix $\bm{A} = (\bm{a}_1, \ldots, \bm{a}_n)^\top$ with
$\bm{a}_i$ in \eqref{eq:ai}, we employ the principal component analysis \citep{bai2003inferential} on the covariance of $\bm{A}$, and further obtain that $\widehat{\bm{U}} = {\bm{A}} - \widehat{\bm{F}}\widehat{\bm{B}}^\top$, where $\widehat{\bm{U}} = \left(\widehat{\bm{u}}_1 ,\ldots ,\widehat{\bm{u}}_n \right)^\top \in \mathbb{R}^{n \times p}$. More specifically, the columns of $\widehat{\bm{F}} / \sqrt{n}$ are the eigenvectors of $\bm{A A}^\top$ corresponding to the top ${K}$ eigenvalues, and $\widehat{\bm{B}}=n^{-1} \bm{A}^\top \widehat{\bm{F}}$. This is similar to the case $\widehat{\bm{B}}=\left(\sqrt{\lambda_1} \bm\xi_1, \ldots, \sqrt{\lambda_{{K}}} \bm\xi_{{K}}\right)$ and $\widehat{\bm{F}}=\bm{A} \widehat{\bm{{B}}} \operatorname{diag}\left(\lambda_1^{-1} \ldots, \lambda_{{K}}^{-1}\right)$, where $\left\{\lambda_j\right\}_{j=1}^{{K}}$ and $\left\{\boldsymbol{\xi}_j\right\}_{i=1}^{{K}}$ are top ${{K}}$ eigenvalues in descending order and their associated eigenvectors of the sample covariance matrix. 

A practical issue is how to determine the number of factors $K$. Given that latent factors, loadings, and idiosyncratic components are all unobservable, a conditional sparsity criterion is then adopted \citep{ahn2013eigenvalue, fan2020factor}. Specifically, let $\lambda_k\left({\bm{A}}^{\top} {\bm{A}}\right)$ denote the $k$-th largest eigenvalue of ${\bm{A}}^{\top} {\bm{A}}$, $K_{\max}$ be a prespecified upper bound for $K$, and $C_n$ be a constant dependent on $n$ and $p=\sum_{g=1}^G m_g$, and then $K$ is determined by 
\begin{equation}\label{Factor_Number}
\widehat{K}=\underset{k \leq K_{\max }}{\operatorname{argmin}} \frac{\lambda_{k+1}\left({\bm{A}}^{\top} {\bm{A}}\right)+C_n}{\lambda_k\left({\bm{A}}^{\top} {\bm{A}}\right)+C_n}
\end{equation}
for a given $C_n$ and $K_{\max}$. An alternative criterion may be the information criteria proposed by \citet{bai2002determining} and \citet{fan2013large}, 
referred to as PC and IC, respectively.

\subsection{Properties of functional factor augmentation structure}
In this subsection, we evaluate the estimation error of the proposed functional factors $\bm{f}_i$. Denote the first $K$ eigenvectors and eigenvalues of $\bm{B}\bm{B}^\top$ as $\bm{\xi}^{(B)}_k$, and $\Lambda_{B} = \operatorname{diag}\{ \lambda_1^{(B)},\ldots,\lambda_K^{(B)}\}$, respectively, where $\lambda_k^{(B)}$, $k = 1, \dots,K$ are sorted in a descending order. The largest eigenvalue of $\operatorname{Cov}(\bm{u}_i)$ is $\lambda_{\max}^{\bm{u}}$, where $\operatorname{Cov}(\bm{u}_i) = \omega\Lambda_u$, the covariance matrix of 
$\bm{a}_i$, $\operatorname{Cov}(\bm{a}_i)$, equals to 
$\mathbb{E}(\bm{a}_i \bm{a}_i^{\top}) = \mathbb{E}(\bm{A}^{\top} \bm{A}/n) = \bm{B} \bm{B}^{\top} + \mathbb{E}( \bm{U}^{\top} \bm{U}/n)$ from \eqref{eq:ai}. The covariance can be further expanded using a perturbation
\begin{align*}
    C_{\bm{a}}(\omega) = \bm{B}\bm{B}^\top + \Lambda_{\bm{u}}\omega + O(\omega^2),
\end{align*}
where $\Lambda_{\bm{u}}$ is a perturbing matrix with $\omega$ such that
\begin{align*}
    \lambda_k^{(C)}(\omega) &= \lambda_k^{(B)} + (\bm{\xi}^{(B)}_k{}^\top \Lambda_{\bm{u}} \bm{\xi}^{(B)}_k)\omega + O(\omega^2),\\
    \bm{\xi}_k^{(C)}(\omega) &= \bm{\xi}^{(B)}_k - (\bm{B}\bm{B}^\top - \lambda_k^{(B)}I)^{+}\Lambda_{\bm{u}}\bm{\xi}_k^{(B)}\omega + O(\omega^2),
\end{align*}
with $\lambda_k^{(C)}(\omega)$ and $\bm{\xi}^{(C)}_k$, $k = 1, \dots, K$ being the first $K$ eigenvalues and eigenvectors of $C_{\bm{a}}(\omega)$ \citep{shi1997local}, respectively. {Based on $C_{\bm{a}}(\omega)$ we have ${\bm{B}(\omega)}=\left(\sqrt{\lambda_1^{(C)}(\omega)} \bm{\xi}_1^{(C)}(\omega), \ldots, \sqrt{\lambda_K^{(C)}(\omega)} \bm{\xi}_K^{(C)}(\omega)\right)$ and ${\bm{F}}(\omega)=\bm{A} \bm{B}(\omega)\left[\operatorname{diag}\left(\lambda_1^{(C)} \ldots, \lambda_K^{(C)}\right)\right]^{-1}$}. Accordingly, the following lemma shows the estimation error of the proposed functional structure factors for a given $\bm{B}$. 
\begin{lemma}\label{lemma:factor}
    For the fixed truncation numbers $m_1,\ldots,m_G$, if $\underset{1 \le i < j \le K}{\min}(\lambda_i^{(B)}-\lambda_j^{(B)}) > c_1 $ and $\lambda_{\max}^{\bm{u}} < c_2$ for some constant $c_1, c_2 > 0$, then
    \begin{align*}
        \Vert{\bm{f}}_i^\top(\omega) - \bm{f}_i^\top\Vert_2^2 &= O_{P}(\omega^2\Vert \bm{a}_i \Vert_2^2 + \Vert \bm{u}_i^\top \bm{B}\Lambda_{B}^{-1} \Vert_2^2),~\quad{\mbox{and}}\\
        \Vert{\bm{u}}_i^\top(\omega) - \bm{u}_i^\top\Vert_2^2 &= O_{P}(\omega^2\Vert \bm{a}_i \Vert_2^2 + \Vert \bm{u}_i^\top \bm{B}\Lambda_{B}^{-1}\bm{B}^\top \Vert_2^2)
    \end{align*}
    with $\bm{f}_i^\top(\omega)$ and $\bm{u}_i^\top(\omega)$ calculated based on $C_{\bm{a}}(\omega)$. 
\end{lemma}

\begin{remark}
    In terms of errors induced by detecting functional structure factors, there are mainly two sources. One is from the estimation error of eigenvalues and eigenvectors, and the other is from the projection of $\bm{u}_i$ in the column space of $\bm{B}$.
\end{remark}

Furthermore, in real applications when only repeated observations $\{(t_{ij}^{(g)},W_{i}^{(g)}(t_{ij}^{(g)})), t_{ij}^{(g)}\in\mathcal{T},j=1,\ldots,n_{i}^{(g)}\}$ are available for each subject, a common practice is to estimate a smooth trajectory $\{\widehat{X}_i^{(g)}(\cdot),i=1,\ldots,n,g=1,\ldots,G\}$ and hence obtain $\widehat{\bm{A}}$, $\widehat{\bm{B}}$ and $\widehat{\bm{F}}$. To further consider the estimation error for $\widehat{\bm{F}}$, more assumptions are required to describe the properties of the scores estimated by fPCA. For each functional covariate, denote the covariance function $\text{Cov}_X^{(g)}(s,t) = \operatorname{Cov}(X_{i}^{(g)}(s),X_{i}^{(g)}(t)),g=1,\ldots,G$, and $\tau_{k}^{(g)}$ is the k-th eigenvalue of covariance functions $\text{Cov}_X^{(g)}(s,t)$. Assume for $g = 1,  \ldots,G$, we have $m_g \leq s_n$.
\begin{assumption}\label{asmp:1}

\begin{itemize}
    \item[(A1)](The decay rates of eigenvalues) For $g = 1,\ldots,G$, $\tau_{1}^{(g)}<\infty$, $\tau_{k}^{(g)}-\tau_{k+1}^{(g)}\geq Ck^{-a-1}$  for $k\geq1$.
    \item[(A2)] (The common truncation parameter $s_n$ cannot be too large) $(s_n^{2a+2}+s_n^{a+4})/n=o(1).$
\end{itemize} 

\end{assumption}

This implies that $\tau_{k}^{(g)}\geq Ck^{-a}$. As the covariance functions $\text{Cov}_X^{(g)}(s,t),g = 1,\ldots,G$ are bounded, one has $a > 1$. 
\begin{assumption} \label{asmp:2}
We defer to the Supplementary Material for Conditions (B1)–(B4) on the underlying processes $X_{i}^{(g)}(t)$. These conditions specify how data are sampled and smoothed.
\end{assumption}

\begin{lemma}\label{lemma:functional}
    For a fixed number $G$ of trajectory types, under Assumptions \ref{asmp:1} and \ref{asmp:2}, 
    \begin{align*}
        \{(\widehat{\bm{A}}^\top\widehat{\bm{A}})_{ij} - E(\bm{A}^\top \bm{A})_{ij}\}/n = O_p(1/\sqrt{n}),\quad \text{for any}~i,j\leq p.
    \end{align*}
    Further, $\Vert \widehat{\bm{a}}_i - {\bm{a}}_i\Vert_2^2 \leq O_P(1/n)$.
\end{lemma}
Lemma \ref{lemma:functional} indicates the convergence rate of estimating $\mathbb{E}(\bm{A}^\top \bm{A})_{ij}/n$ using $(\widehat{\bm{A}}^\top\widehat{\bm{A}})_{ij}$ is the same as using $({\bm{A}}^\top {\bm{A}})_{ij}$. Combined with Lemma \ref{lemma:factor}, the estimation error of functional structure factors is formally established.

\begin{theorem}\label{thm:factor-bound}
    With $\widehat{\bm{F}}=\widehat{\bm{A}} \widehat{\bm{{B}}} \operatorname{diag}\left(\widehat{\lambda}_1^{-1} \ldots, \widehat{\lambda}_{{K}}^{-1}\right)$ and $\widehat{\bm{U}} = \widehat{\bm{A}} - \widehat{\bm{F}}\widehat{\bm{B}}^\top$, we have
    \begin{align*}
    \left\{
    \begin{array}{l}
    \Vert\widehat{\bm{f}}_i^\top - \bm{f}_i^\top\Vert_2^2 = O_P((\omega^2+ 1/n)\Vert \bm{a}_i \Vert_2^2 + \Vert \bm{u}_i^\top \bm{B}\Lambda_{B}^{-1} \Vert_2^2 ),\\
        \Vert\widehat{\bm{u}}_i^\top - \bm{u}_i^\top\Vert_2^2 = O_P((\omega^2+ 1/n)\Vert \bm{a}_i \Vert_2^2  + \Vert \bm{u}_i^\top \bm{B}\Lambda_{B}^{-1}\bm{B}^\top \Vert_2^2 ).
    \end{array}\right.
    \end{align*}
\end{theorem}
    
\subsection{Truncation analysis and the relationship between $K$ and $m_g$}

A practical issue when modeling functional data with sample instances is how to determine the number of basis functions using truncation, i.e., $m_g$ for $X_i^{(g)}(t)$, and how this truncation would affect the estimate of functional factors $\bm{f}_i$ in \eqref{eq:factor}. Usually, $m_g$ would be determined by the cumulative contribution to the variance from the corresponding functional scores, which tends to get an overestimated value of $m_g$, namely, $\widehat{m}_g > m_g$. To illustrate, assume $\widehat{m}_1 > m_1$ and $\widehat{m}_g = m_g$ for $g=2,\ldots,G$, and define $\tilde{\bm{a}}_{i} = (\bm{a}_{i}^{(1)\top},\tilde{\bm{a}}_{i}^{(1) \top},\bm{a}_{i}^{(2)\top},\ldots,\bm{a}_{i}^{(G)\top})^{\top}$ where $\tilde{\bm{a}}_{i}^{(g) \top}$ is a redundant variable for the factor model. In this case, 
 
\begin{align*}
    \operatorname{Cov}((\bm{a}_{i}^{(g) \top},\tilde{\bm{a}}_{i}^{(g) \top})^\top ) = \begin{pmatrix}
        \Sigma_{m_g} & \bm{0}\\
        \bm{0} & \tilde{\Lambda}_0^{(g)}
    \end{pmatrix},
\end{align*}
and
\begin{align*}
    \operatorname{Cov}(\tilde{\bm{a}}_i) = \begin{pmatrix}
 \Lambda_{m_1} & \bm{0} & \Sigma_{m_1m_2} & \cdots  & \Sigma_{m_1m_G}\\
 \bm{0} & \tilde{\Lambda}_0^{(1)} & \bm{0} &  \cdots  & \bm{0}\\
  \Sigma_{m_2m_1}& \bm{0} &  \Lambda_{m_2}& \cdots & \Sigma_{m_2m_G}\\
  \vdots& \vdots & \vdots &  \ddots& \vdots\\
  \Sigma_{m_Gm_1}& \bm{0} & \Sigma_{m_Gm_2} & \cdots &\Lambda_{m_G}
\end{pmatrix}.
\end{align*}
If $\xi^{\star} = (\xi_1^{\star\top},\xi_2^{\star\top})^\top$ is one of the first $K$ {eigenvectors} of $\operatorname{Cov}({\bm{a}}_i)$, $(\xi_1^{\star\top},\bm{0},\xi_2^{\star\top})^\top$ is an eigenvector of $\operatorname{Cov}(\tilde{\bm{a}}_i)$ with the same eigenvalue, and the position of $\bm{0}$ in $(\xi_1^{\star\top},\bm{0},\xi_2^{\star\top})^\top$ corresponds to that of $\tilde{\Lambda}_0^{(1)}$ in $\operatorname{Cov}(\tilde{\bm{a}}_i)$. As $\widehat{\bm{F}}=\bm{A} \widehat{\bm{B}} \operatorname{diag}\left(\lambda_1^{-1} \ldots, \lambda_K^{-1}\right)$, $\tilde{\bm{a}}_{i}^{(g)}$ will not affect $\widehat{\bm{F}}$ since the row of $\bm{B}$ corresponding to $\tilde{\bm{a}}_{i}^{(g)}$ is $\bm{0}$ when ${\widehat{K}} = K$. Actually, in this overestimating situation, we can treat 
\begin{align*}
\tilde{\bm{a}}_{i}
 = \begin{pmatrix}
         B_1  \\
         \bm{0}\\
         B_2 
    \end{pmatrix}\bm{f}_i + \begin{pmatrix}
         \bm{u}_i^{(1)}  \\
         \tilde{\bm{a}}_{i}^{(1)}\\
         \bm{u}_i^{(C)} 
    \end{pmatrix} = B^\star\bm{f}_i+\tilde{\bm{u}}_{i},
\end{align*}
as the real model with $\bm{u}_i = (\bm{u}_i^{(1)}{}^\top,\bm{u}_i^{(C)}{}^\top)^\top$ and $\bm{B} = \left(B_1^\top,B_2^\top\right)^\top$.

Another case is that $m_{g}$ may be underestimated for some specific $g_0\leq G$, especially when $\omega$ and $K$ are relatively small. In this case, a majority amount of variance of $X_i(t)$ is concentrated in ${{\tilde{\bm{\phi}_0}}{}^{(g_0)}(t)}{}^{\top}\bm{f}_i$ by \eqref{eq:basis3}. When \(\tilde{\bm{\phi}}_0^{(g_0)}(t)\) are linearly independent for $g=1,\ldots,G$, $\widehat{m}_{g_0}$ is determined as $K$ by the variance contribution criteria. However, when they are correlated, an underestimation of $m_{g_0}$ may not significantly affect the estimation of $\widehat{\bm{f}}_i$. To illustrate, consider that the covariance matrix of $\bm{a}_{i}$ with $G=2$ as
\begin{align*}
    {\operatorname{Cov}(\bm{a}_{i})} = \begin{pmatrix}
 \Lambda_{m_1} & \Sigma_{m_1m_2}\\
  \Sigma_{m_2m_1}&  \Lambda_{m_2}
\end{pmatrix},
\end{align*}
with $K<m_1,m_2$, and for convenience, we assume $\Lambda_u = \bm{I}_p$. Since $B^{(g)}$ is an $m_g \times K$ matrix, $B^{(g)}B^{(g)}{}^\top$ has at most $K$ non-zero eigenvalues, denoted as $\nu_j^{(g)},j=1,\ldots,K$. By plugging in $\Lambda_{m_g} = P^{(g)}{}^\top B^{(g)}B^{(g)}{}^\top P^{(g)} + \omega \bm{I}_{m_g} = \operatorname{diag}\{\nu_1^{(g)} + \omega,\ldots,\nu_K^{(g)} + \omega,\omega,\ldots,\omega\}$, ${\operatorname{Var}(\bm{a}_{i})}$ is written as
\begin{align*} \label{eq:underestimate}
   {\operatorname{Cov}(\bm{a}_{i})} &= \begin{pmatrix}
 \Lambda_{m_{1(K)}}& 0 & \Sigma_{m_{1(K)}m_{2(K)}} &\Sigma_{m_{1(K)}m_{2(C)}}\\
 0 & \Lambda_{m_{1(C)}} & \Sigma_{m_{1(C)}m_{2(K)}} &\Sigma_{m_{1(C)}m_{2(C)}}
 \\
  \Sigma_{m_{2(K)}m_{1(K)}} &\Sigma_{m_{2(K)}m_{1(C)}} & \Lambda_{m_{2(K)}}& 0 \\
  \Sigma_{m_{2(C)}m_{1(K)}} &\Sigma_{m_{2(C)}m_{1(C)}} &
  0 & \Lambda_{m_{2(C)}}
\end{pmatrix} \\
&\approx 
\begin{pmatrix}
 \Lambda_{m_{1(K)}}& 0 & \Sigma_{m_{1(K)}m_{2(K)}} &0\\
 0 & 0 & 0 & 0
 \\
  \Sigma_{m_{2(K)}m_{1(K)}} & 0 & \Lambda_{m_{2(K)}}& 0 \\
  0 & 0 &
  0 & 0
\end{pmatrix}, 
\end{align*}
with $\Lambda_{m_g(K)} = \operatorname{diag}\{\tau_1^{(g)} + \omega,\ldots,\tau_K^{(g)} + \omega\}$ and $\Lambda_{m_g(C)} = \operatorname{diag}\{\omega,\ldots,\omega\}$. Consequently, when $\omega \approx 0$, underestimating $m_g$ will not significantly affect $\widehat{\bm{f}}_i$ as long as $\widehat{m}_g \geq K$, which is easy to achieve in practice.

\section{The functional factor augmentation selection model }\label{sec:fFASM}

\subsection{The functional linear regression model}
In this section, we address a multivariate functional linear regression model with correlated functional covariates, using the proposed fFAS in Section 2. To start with, consider a functional linear regression model where a scalar response $Y$ with $\mathbb{E}(Y) = \mu_Y$ is generated by a group of $G$ functional covariates $\{X^{(1)}(t), X^{(2)}(t),\ldots, X^{(G)}(t); t\in \mathcal{T}\}$ as
\begin{equation}\label{eq:flr_pop}
    Y = \mu_Y + \sum_{g = 1}^G\int_{\mathcal{T}} \beta^*{}^{(g)}(t) \left(X^{(g)}(t)-\mu^{(g)}(t)\right)dt + \epsilon,
\end{equation}
where $\beta^*{}^{(g)}(\cdot)$ is the square-integrable regression parameter function, and $\epsilon$ is a random noise with a zero mean and a constant variance $\sigma^2$. We note $\beta_0^* = \mu_Y$ as the intercept term. Thus, with given i.i.d samples $\{y_i, X_i^{(g)}(t),g = 1, \ldots,G; t\in \mathcal{T}, i=1,\ldots,n$\} which have the same distribution as $\{Y, X^{(g)}(t),g = 1, \ldots,G; t\in \mathcal{T}\}$, the sample functional linear regression model is described as
\begin{align}\label{eq:flr_sample}
    y_i &= \beta_0^* + \sum_{g = 1}^G\int_{\mathcal{T}} \beta^*{}^{(g)}(t) \left(X_i^{(g)}(t) - \mu^{(g)}(t)\right)dt + \epsilon_i.
\end{align}

To detect active functional covariates with correlation, we develop a functional factor augmentation selection model (fFASM) as follows. Without loss of generalizability, we use $X_i^{(g)}(\cdot)$ and $y_i$ as the centered functional covariates and scalar response variable, respectively, and accordingly \eqref{eq:flr_sample} is equivalent to 
\begin{equation}\label{eq:linear1}
    y_i = \sum_{g = 1}^G\int_{\mathcal{T}} \beta^*{}^{(g)}(t) X_i^{(g)}(t)dt + \epsilon_i,
\end{equation}
and can be further expanded by K-L expansion as
\begin{align*}
    y_i &= \sum_{g = 1}^G\sum_{j=1}^{m_g}a_{ij}^{(g)}\eta_{j}^*{}^{(g)} + \sum_{g = 1}^G\int_{\mathcal{T}} \beta^*{}^{(g)}(t) e_i^{(g)}(t)dt + \epsilon_i = \bm{H}^*{}^\top \cdot\bm{a}_i \ + \tilde{\epsilon}_i,
\end{align*}
where $\bm{H}^* = \left(\bm{\eta}^*{}^{(1)}{}^\top,\ldots,\bm{\eta}^*{}^{(G)}{}^\top\right)^\top \in \mathbb{R}^{p},\eta_j^*{}^{(g)} = \int_{\mathcal{T}}\beta^*{}^{(g)}(t){\gamma}_j^{(g)}(t)dt, j=1,\ldots,m_g$, and $\tilde{\epsilon}_i = \epsilon_i + \sum_{g = 1}^G\int_{\mathcal{T}} \beta^*{}^{(g)}(t) e_i^{(g)}(t)dt$. For $g = 1,\ldots,G$, $\int_{\mathcal{T}} \beta^*{}^{(g)}(t) e_i^{(g)}(t)dt$ has a zero mean. Consequently, to select useful functional covariates $X_i^{(g)}(t)$ is to find $\beta^*{}^{(g)}(\cdot)$ such that $\beta^*{}^{(g)}(\cdot) \neq 0$, which is further assumed as $\bm{\eta}^*{}^{(g)} \neq \bm{0}$. Plugging in the proposed fFAS, one easily obtains
\begin{align*}
    y_i = \bm{H}^*{}^{\top}\left(\bm{B}\bm{f}_i + \bm{u}_i\right) + \tilde{\epsilon}_i,
\end{align*}
or equivalently,
\begin{align}\label{eq:lr factor}
    \bm{y} = \bm{F}\bm{B}^{\top} \bm{H}^*{} + \bm{U} \bm{H}^*{} + \tilde{\bm{\epsilon}},
\end{align}
and with $\widehat{\bm{F}}$, $\widehat{\bm{B}}$ and $\widehat{\bm{U}}$ obtained by the fFAS, \eqref{eq:lr factor} is further equivalent to 
\begin{align}\label{eq:proj}
    \left(\bm{I}_n - \bm{P}_{\widehat{\bm{F}}}\right)\bm{y} &= \left(\bm{I}_n-\bm{P}_{\widehat{\bm{F}}}\right)\widehat{\bm{U}} \bm{H}^*{} + \left(\bm{I}_n-\bm{P}_{\widehat{\bm{F}}}\right)\left(\tilde{\bm{e}} + \tilde{\bm{\epsilon}} \right),
\end{align}
where $\bm{P}_{\widehat{\bm{F}}} = {\widehat{\bm{F}}}({\widehat{\bm{F}}}^\top {\widehat{\bm{F}}})^{-1}{\widehat{\bm{F}}}^\top$ is the orthogonal projection matrix onto the column space $C({\widehat{\bm{F}}})$, and $\tilde{\bm{e}} = (\tilde{e}_1,\ldots,\tilde{e}_n)^\top = \left(\bm{A} - \widehat{\bm{A}} + \mathbb{E}(\widehat{\bm{A}})\right)\bm{H}^*{}$. Consequently to select useful functional covariates, the penalized loss function 
\begin{align*}
L_n\left(\bm{y},\widehat{\bm{U}}\bm{H}, \widehat{\bm{F}}\right) = \frac{1}{n}\left\|(\bm I - P_{\widehat{\bm{F}}})(\bm{y} - \widehat{\bm{U}} \bm{H}) \right\|_2^2 + \sum_{g=1}^{G}\sum_{k=1}^{m_g}J_{\lambda}\left(\eta_{k}^{(g)}\right)
\end{align*}
is minimized with respect to $\bm{H}$, where $J_{\lambda} = \lambda \cdot J(\cdot)$ is a penalty controlled by the parameter $\lambda$ and $J(\cdot)$ can be set as  the popular penalties such as lasso, SCAD or MCP. Note that \(\lambda\) can be selected using cross-validation. Hence, we successfully transform the problem from model selection with highly correlated functional covariates to model selection with weakly correlated or uncorrelated ones by lifting the space to higher dimension. As $\widehat{\bm{H}} = \argmin_{\bm{H}} L_n(\bm{y},\widehat{\bm{U}}\bm{H}, \widehat{\bm{F}})$ is obtained, ${\beta}^{(g)}(t)$ is estimated as
\begin{align*}
    \widehat{\beta}^{(g)}(t) = \sum_{j=1}^{\widehat{m}_g} \widehat{\eta}_j^{(g)}\widehat{\gamma}_j^{(g)}(t).
\end{align*}
Accordingly, when $\widehat{\beta}^{(g)}(t) \neq 0$, ${X_i}^{(g)}(t)$ is selected as a useful functional covariate. Note that the group selection method, such as the GM strategy by \citet{ANEIROS2022104871}, is not adopted in our case.

Also, this procedure may work even in the generalized linear context. Honestly, 
\begin{align} \label{eq:linear}
    \bm{y} &= \left( \widehat{\bm{A}} - \mathbb{E}(\widehat{\bm{A}}) \right)\bm{H}^* + \left(\bm{A} -  \widehat{\bm{A}} + \mathbb{E}(\widehat{\bm{A}}) \right)\bm{H}^*{}+\tilde{\bm{\epsilon}}\notag\\
    &= \widehat{\bm{F}}\widehat{\bm{B}}^{\top}\bm{H}^*{} + \widehat{\bm{U}} \bm{H}^*{} + \left(\bm{A} -  \widehat{\bm{A}} + \mathbb{E}(\widehat{\bm{A}}) \right)\bm{H}^*{}+\tilde{\bm{\epsilon}}\notag\\
    &= \widehat{\bm{F}}\widehat{\bm{B}}^{\top}\bm{H}^*{} + \widehat{\bm{U}}\bm{H}^*{} + \tilde{\bm{e}} + \tilde{\bm{\epsilon}},
\end{align}
indicating that the explanatory variables are switched from $\bm{A}$ to $\widehat{\bm{A}}$. In practice, when using the sample mean of $\widehat{\bm{A}}$ to substitute $\mathbb{E}(\widehat{\bm{A}})$,  $\mathbb{E}(\widehat{\bm{a}}_i)\rightarrow 0$ under some regularization conditions \citep{kong2016partially}. After centralizing $\widehat{\bm{A}}$, by $\left( \widehat{\bm{A}} - \mathbb{E}(\widehat{\bm{A}}) \right)\bm{H}^*{} = \widehat{\bm{F}}\widehat{\bm{B}}^{\top}\bm{H}^*{} + \widehat{\bm{U}}\bm{H}^*{} = \widehat{\bm{F}}\bm{\gamma}^* + \widehat{\bm{U}}\bm{H}^*{}$, the unknown parameters are transformed into $(\bm{H},\bm{\gamma})$, so that their corresponding covariates $\widehat{\bm{U}}$ and $\widehat{\bm{F}}$ are weakly correlated by introducing $\bm \gamma$. Note that in linear cases, $\bm{\gamma}^*$ is further eliminated by using the projection matrix in \eqref{eq:proj}. Consequently, for generalized linear models and samples without centralization, the loss function is updated as
\begin{align}\label{loss:g}
L_n\left(\bm{y},\widehat{\bm{W}}\bm{\theta}\right) =& \frac{1}{n}\sum_{i=1}^n\left[ -y_i\left(\beta_0 + \widehat{\bm{f}}_i^\top \bm{\gamma} + \widehat{\bm{u}}_i^\top \bm{H}\right) + b\left(\beta_0 + \widehat{\bm{f}}_i^\top \bm{\gamma} + \widehat{\bm{u}}_i^\top \bm{H}\right) \right]\nonumber \\
&+ \sum_{g=1}^{G}\sum_{k=1}^{m_g}J_{\lambda}\left(\eta_{k}^{(g)}\right) + J_\lambda(\beta_0),
\end{align}
where $\widehat{\bm{w}}_i = \left(1,  \widehat{\bm{u}}_i^\top,\widehat{\bm{f}}_i^\top\right)^\top$,
$\widehat{\bm{W}} = \left(  \widehat{\bm{w}}_1, \ldots, \widehat{\bm{w}}_n \right)$
and $\bm{\theta} = \left(\beta_0,  \bm{H}^\top,\bm{\gamma}^\top\right)^\top$, $\beta_0$ is the intercept term and $b(\cdot)$ is a known function, where $b(z) = z^2/2$ in linear models.  The estimate is given by $\left(\widehat{\beta_0}, \widehat{\bm{H}}^\top, \widehat{\bm{\gamma}}^\top \right)^\top = \argmin_{\bm{\theta}}L_n\left(\bm{y},\widehat{\bm{W}}\bm{\theta}\right)$.

\subsection{Theoretical justifications for functional variable selection}
In this section, the proposed method is theoretically investigated under the general linear model context in \eqref{loss:g}, and hence the linear model will be covered as a special case. To start with, some notations and assumptions are introduced. Recall $p=\sum_{g=1}^G m_g$, and define $p_1 = 1 + p $, $\bm{H}_1 = (\beta_0,\bm{H}^\top)^\top$, $\bm{H}_1^* = (\beta_0^*,\bm{H}^*{}^\top)^\top$, $\bm{\theta}^* = (\beta_0^*,  \bm{H}^*{}^\top,(\bm{B}^\top\bm{H}^*)^\top)^\top$, $S = \operatorname{supp}(\bm{\theta}^*)$, $S_1 = \operatorname{supp}(\bm{H}_1^*)$, and $S_2 = [p_1+K] \backslash S$. Suppose $\widehat{\bm{F}}$ and $\widehat{\bm{U}}$ are obtained given $\bm{A}$. We write
\begin{align}
    X_i^{(g)}(t) &=  X_{f,i}^{(g)}(t) + X_{u,i}^{(g)}(t) + e_{0i}^{(g)}(t),
\end{align}
with $X_{f,i}^{(g)}(t) = \bm{\gamma}^{(g)}(t){}^{\top} B^{(g)}\bm{f}_i$ and $X_{u,i}^{(g)}(t) = {\bm{\gamma}^{(g)}(t)}{}^{\top} \bm{u}_i^{(g)}$.

\begin{assumption} \label{asmp:sm}
    (Smoothness). $b(z) \in C^3(\mathbb{R})$, i.e., for some constants $M_2$ and $M_3$, $0 \leq b^{\prime \prime}(z) \leq M_2$ and $\left|b^{\prime \prime \prime}(z)\right|$ $\leq M_3, \forall z$.
\end{assumption}
\begin{assumption} \label{asmp:convexity irrepresentable}
    (Restricted strong convexity and irrepresentable condition). For ${\bm{W}} = \left(  {\bm{w}}_1, \ldots, {\bm{w}}_n \right)$ where ${\bm{w}}_i = \left(1,  {\bm{u}}_i^\top,{\bm{f}}_i^\top\right)^\top$, there exist $\kappa_2>\kappa_{\infty}>0$ and $\tau \in(0,0.5)$ such that 
    \begin{align}
    &\text{(Convexity)}    \left\|\left[\nabla_{S S}^2 L_n\left(\bm{y}, \bm{W} \boldsymbol{\theta}^*\right)\right]^{-1}\right\|_{\ell} \leq \frac{1}{4 \kappa_{\ell}}, \quad \text { for } \ell=2 \text { and } \infty, \label{eq:asmp:1}\\
    &\text{(Irrepresentable condition)}  \left\|\nabla_{S_2 S}^2 L_n\left(\bm{y}, \bm{W} \boldsymbol{\theta}^*\right)\left[\nabla_{S S}^2 L_n\left(\bm{y}, \bm{W} \boldsymbol{\theta}^*\right)\right]^{-1}\right\|_{\infty} \leq 1-2 \tau .\label{eq:asmp:2}
    \end{align}

\end{assumption}

\begin{assumption}\label{asmp:estm of fm}
    (Estimation of factor model). $\Vert \gamma_j^{(g)}(\cdot) \Vert_{\infty}\leq M_{\gamma}$ for $j = 1,\ldots,m_g,~g = 1,\ldots,G$, and $\Vert X_{f,i}^{(g)}(\cdot)\Vert_\infty \leq \frac{1}{2}M_{\gamma}M_0 \sqrt{\lambda_1^{(B)}}  $ and $\Vert X_{u,i}^{(g)}(\cdot)\Vert_\infty \leq  \frac{1}{2}M_{\gamma}M_0$ for some constant $M_0 > 2$. In addition, there exists a $K \times K$ nonsingular matrix $\bm{V}_0$, 
    and $\bm{V}=\left(\begin{array}{cc}\bm{I}_{p_1} & \bm{0}_{{p_1} \times K} \\ \bm{0}_{K \times {p_1}} & \bm{V}_0\end{array}\right)$ such that for $\overline{\bm{W}}=\widehat{\bm{W}} \bm{V}$, we have $\|\overline{\bm{W}}-\bm{W}\|_{\max } \leq \frac{M_0}{2}$ and $\max _{j \in[p_1+K]}$ $\left(\frac{1}{n} \sum_{i=1}^n\left|\bar{w}_{i j}-w_{i j}\right|^2\right)^{1 / 2} \leq \frac{2 \kappa_{\infty} \tau}{3 M_0 M_2|S|}$, with $w_{i j}$ and $\bar{w}_{i j}$ being the $(i, j)$-th element of $\mathbf{W}$ and $\overline{\mathbf{W}}$. 
\end{assumption}

Assumption \ref{asmp:sm} holds for a large family of generalized linear models. For example, linear model has $b(z)=\frac{1}{2} z^2$, $M_2=1$, and $M_3=0$; logistic model has $b(z)=\log \left(1+e^z\right)$ and finite $M_2, M_3$. Assumption 4 is easily satisfied with a small matrix and holds with high probability as long as $\mathbb{E}\left[\nabla^2 L_n\left(\bm{y}, \bm{W} \boldsymbol{\theta}^*\right)\right]$ satisfies similar conditions by standard concentration inequalities \citep{merlevede2011bernstein}. Assumption \ref{asmp:estm of fm}
indicates the norm $\|\bm{W}\|_{\max }$ when $ X_i^{(g)}(\cdot)$ and $\varepsilon_i^{(g)}(\cdot)$ is bounded, and 
\begin{align*}
    \Vert\widehat{\bm{f}}_i^\top - \bm{f}_i^\top\Vert_2^2 &= O(\omega^2 + \Vert \bm{u}_i^\top \bm{B}\Lambda_{B}^{-1} \Vert_2^2 + 1/n),\\
    \Vert\widehat{\bm{u}}_i^\top - \bm{u}_i^\top\Vert_2^2 &= O(\omega^2  + \Vert \bm{u}_i^\top \bm{B}\Lambda_{B}^{-1}\bm{B}^\top \Vert_2^2 + 1/n),
\end{align*} 
will be satisfied with a high probability when $n$ is large and $\omega$ is small. Furthermore, with $J(\cdot) = |\cdot|$ and $\bm{\widehat{H}}$ obtained by minimizing \eqref{loss:g}, the following result is established.

\begin{theorem}\label{thm1}
    Suppose Assumptions \ref{asmp:sm}-\ref{asmp:estm of fm} holds. Define $M=M_0^3 M_3|S|^{3 / 2}$, and
$$
\varepsilon^*=\max _{j\in[p_1+K]}\left|\frac{1}{n} \sum_{i=1}^n \bar{w}_{i j}\left[-y_i+b^{\prime}\left( {\bm{a}}_i^T \bm{H}^* + \beta_0^*\right)\right]\right| .
$$
If $\frac{7 \varepsilon^*}{\tau}<\lambda<\frac{\kappa_2 \kappa_{\infty} \tau}{12 M \sqrt{|S|}}$, then $\operatorname{supp}(\widehat{\bm{H}}) \subseteq \operatorname{supp}\left(\bm{H}^*\right)$ and
$$
\left\|\widehat{\bm{H}}-\bm{H}^*\right\|_{\infty} \leq \frac{6 \lambda}{5 \kappa_{\infty}}, \quad\left\|\widehat{\bm{H}}-\bm{H}^*\right\|_2 \leq \frac{4 \lambda \sqrt{|S|}}{\kappa_2}, \quad\left\|\widehat{\bm{H}}-\bm{H}^*\right\|_1 \leq \frac{6 \lambda|S|}{5 \kappa_{\infty}} .
$$
In addition, if $\varepsilon^*<\frac{\kappa_2 \kappa_{\infty} \tau^2}{12 C M \sqrt{|S|}}$ and $\min \left\{\left|{\bm{H}_1^*}_j\right|: {\bm{H}_1^*}_j \neq 0, j \in[p_1]\right\}>\frac{6 C \varepsilon^*}{5 \kappa_{\infty} \tau}$ hold for some $C>7$, where ${\bm{H}_1^*}_j$ means the j-th element of ${\bm{H}_1^*}$, then by taking $\lambda \in\left(\frac{7}{\tau} \varepsilon^*, \frac{C}{\tau} \varepsilon^*\right)$ the sign consistency $\operatorname{sign}(\widehat{\bm{H}})=\operatorname{sign}\left(\bm{H}^*\right)$ is achieved.
\end{theorem}

Theorem \ref{thm1} guarantees the selection consistency of the functional covariates by developing sign consistency under some mild conditions. Furthermore, it demonstrates the relationship between $\left\|\widehat{\bm{H}}-\bm{H}^*\right\|$ and $\lambda$ whose value depends on $\varepsilon^*$ and $\tau$, where $\varepsilon^*$ comes from the first-order partial derivative of the empirical loss function (without penalty) when the parameters are known and $\tau$ satisfying a generalized irrepresentable condition from \eqref{eq:asmp:2} \citep{lee2015model}. When $\varepsilon^*$ is small, selecting an appropriate $\lambda$ leads to satisfactory performance in the model estimation. Furthermore, when $\bm{A}$ is not available, $\widehat{\bm{A}}$ will be estimated and employed. Accordingly, the sign consistency is further guaranteed by showing that the loss function \eqref{loss:g} based on $\widehat{\bm{A}}$ will have an asymptotic property as follows.
\begin{theorem}
    Let $L_n\left(\bm{y},\widehat{\bm{W}}\bm{\theta}\right)|_{\bm{A}}$ be the loss function in \eqref{loss:g}, based on which $\widehat{\bm{f}}_i$ and $\widehat{\bm{u}}_i$ are obtained using the true $\bm{A}$, and $L_n\left(\bm{y},\widehat{\bm{W}}\bm{\theta}\right)|_{\widehat{\bm{A}}}$ using ${\widehat{\bm{A}}}$. Then for any fixed $\bm{\theta}$, by Assumption \ref{asmp:sm} and Lemma \ref{lemma:functional}, 
    \begin{align*}
    \left|L_n\left(\bm{y},\widehat{\bm{W}}\bm{\theta}\right)|_{\bm{A}} - L_n\left(\bm{y},\widehat{\bm{W}}\bm{\theta}\right)|_{\widehat{\bm{A}}} \right| \rightarrow 0 \text{ in probability}.
    \end{align*}
\end{theorem}

\begin{corollary}
    Suppose $L_n\left(\bm{y},\widehat{\bm{W}}\bm{\theta}\right)$ has a unique global minimum. Then $L_n\left(\bm{y},\widehat{\bm{W}}\bm{\theta}\right)|_{\bm{A}}$ and $L_n\left(\bm{y},\widehat{\bm{W}}\bm{\theta}\right)|_{\widehat{\bm{A}}}$ will have an identical global minimizer when $n$ is sufficiently large, so that the estimates from $L_n\left(\bm{y},\widehat{\bm{W}}\bm{\theta}\right)|_{\widehat{\bm{A}}}$ will also have sign consistency.
\end{corollary}

\section{Simulation Studies}\label{sec:Simulation}
In this section, the performance of the proposed fFASM estimator is examined for correlated functional data using simulation studies. To start with, the functional covariates $\{X_i^{(g)}(t)\}_{g=1}^G$ are generated by $X_i^{(g)}(t) =\bm{a}_{0,i}^{(g)}{}^\top \cdot \bm{\phi}_{0,m_g}^{(g)}(t) + \varepsilon_i(t)$ with $\varepsilon_i(\cdot) $ independently and identically (i.i.d) drawn from $N(0,0.25)$, and sampled over 51 uniformly distributed grid points for $t$ from $\mathcal{T}=[0,1]$ with interval lengths of 0.02, where $G$ is set as 20, 50, 100 and 150, respectively. The basis functions $\bm{\phi}_{0,m_g}^{(g)}(t) = \bm{\phi}_{m}(t) = (\phi_1(t),\ldots,\phi_{m}(t))^{\top}$ are Fourier basis with $m$ set as 10. To blend correlations into functional covariates $X_{i}^{(g)}(t)$, two scenarios are considered when $\bm{a}_i$ is generated:
    \begin{itemize}
    \item Scenario I: the factor model case, i.e.,
    $\bm{a}_{0,i} = \left(\bm{a}_{0,i}^{(1)\top}{},\ldots,\bm{a}_{0,i}^{(G)\top}{}\right)^\top = \bm{B}\bm{f}_i + \bm{u}_i,$ where the elements of $\bm{f}_i \in \mathbb{R}^{K}$, $\bm{u}_i \in \mathbb{R}^{Gm}$ and $\bm{B} \in \mathbb{R}^{Gm \times K}$  are generated from  $N(0, 25)$, $N(0, 1)$, and $N(0,1)$, respectively. The true number of factors $K$ is set as $1,2,\ldots,6$, respectively;
    \item Scenario II: the equal correlation case, i.e., $\bm{a}_{0,i}$ are generated from a multivariate normal distribution $\bm{a}_{0,i} \sim N(\bm{0}, \Sigma_{Equal})$
    where $\Sigma_{Equal}$ has diagonal elements as 1 and off-diagonal elements as $\rho$. The true value of $\rho$ is set as $0, 0.1, 0.2 \ldots,0.9$, respectively.
\end{itemize} 
Furthermore, the true functional coefficients $\beta^{*(g)}$ in \eqref{eq:linear1} are generated 
$\beta^{*(1)}(t) = \beta^{*(2)}(t) = \beta^{*(3)}(t) = \beta^{*(4)}(t) = (1, 1/2^2, \ldots, 1/m^2) \cdot \bm{\phi}_{m}(t)$ (the Harmonic attenuation type), $\beta^{*(5)}(t) = \beta^{*(6)}(t) = (0, 1/2^2, 0, \ldots, 0) \cdot \bm{\phi}_{m}(t)$ (the weak single signal type), and $\beta^{*(g)}(t)=0$ for $g=7,8,\ldots,G$. Accordingly, the response $y_i$ is generated by \eqref{eq:linear1} with $\epsilon_i \sim N(0,0.1)$. 

We compare the performance of the proposed fFASM method (MCP penalty) with other functional variable selection methods, including the MCP (\texttt{MCP}) and the group MCP (\texttt{grMCP}) methods on the synthetic data in different scenarios. To evaluate the performance, three different measurements are adopted, i.e, the model size defined as the cardinality of the set $\left\{g : \widehat{\beta}^{(g)}(t) \neq 0\right\}$, the integrated mean squared error (IMSE) 

$$
IMSE = \sum_{g=1}^G \mathbb{E}\left(\left\|\widehat{\beta}^{(g)}(t) - \beta^{*(g)}(t)  \right\|^2_{L^2}\right)
=  \sum_{g=1}^G \mathbb{E} \int_\mathcal{T} \left(\widehat{\beta}^{(g)}(t) - \beta^{*(g)}(t)\right)^2 dt,
$$
and the true positive rate (TPR)

$$
\text{TPR} = \frac{\text{TP}}{\text{TP} + \text{FN}},
$$
where TP represents the number of correctly predicted nonzero instances of
$\widehat{\beta}^{(g)}(t)$ (events that are actually positive and predicted as positive), 
and FN for that of instances that are actually nonzero but are predicted as zero. Note that a model with the TPR of 1 and the model size of 6 indicates perfect recovery. The proposed fFASM method employs the MCP regularization, and the hyperparameters are tuned using cross-validation by minimizing the cross-validation error with a randomly selected subset of $[n/3]$ training samples, where the sample size $n$ is set as 100. The whole experiment is repeated for 500 times, and the averaged results are reported.

\begin{figure}[h] 
    \centering
    \includegraphics[width=1\textwidth]{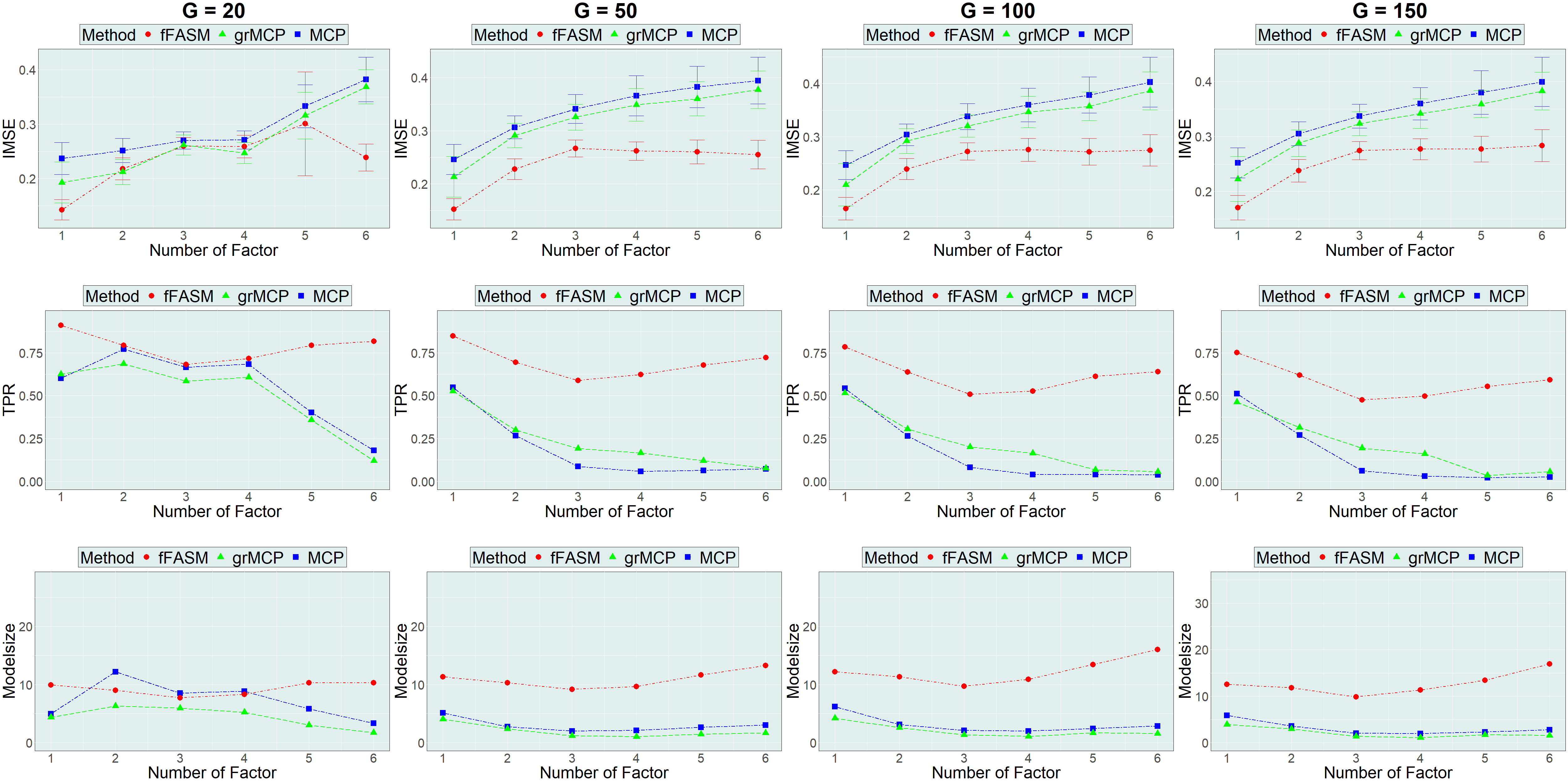}
    \caption{Estimation and selection performance of three methods with different numbers of factors $K$ for Scenario I. In each row, four subfigures are displayed with $G =20,50,100,150$, respectively, and in each column with a given $K$, the IMSE (with 0.5 standard deviations), the TPR, and model size are presented, respectively.}
    \label{fig1}
\end{figure}

Figure \ref{fig1} shows the estimation and selection performance for Scenario I (the factor model structure case) using different methods. As is easily observed, the proposed fFASM method significantly and consistently outperforms MCP and grMCP in the sense that the proposed fFASM has the smallest IMSE for each $K$, demonstrating a more accurate estimation of the functional coefficients. Furthermore, the TPR for the proposed fFASM method is always greater than those of MCP and grMCP as with the increase of $K$, and MCP and grMCP even show dramatic drops when $K$ increases. In terms of the model size, the proposed fFASM tends to select more functional covariates into the model, slightly greater than the true model size 6, while the MCP and grMCP methods tend to select irrelevant functional covariates into the model when looking deeper at the selection results, and the grMCP method tends to be more conservative in selecting functional covariates than MCP.  

\begin{figure}[h] 
    \centering
    \includegraphics[width=1\textwidth]{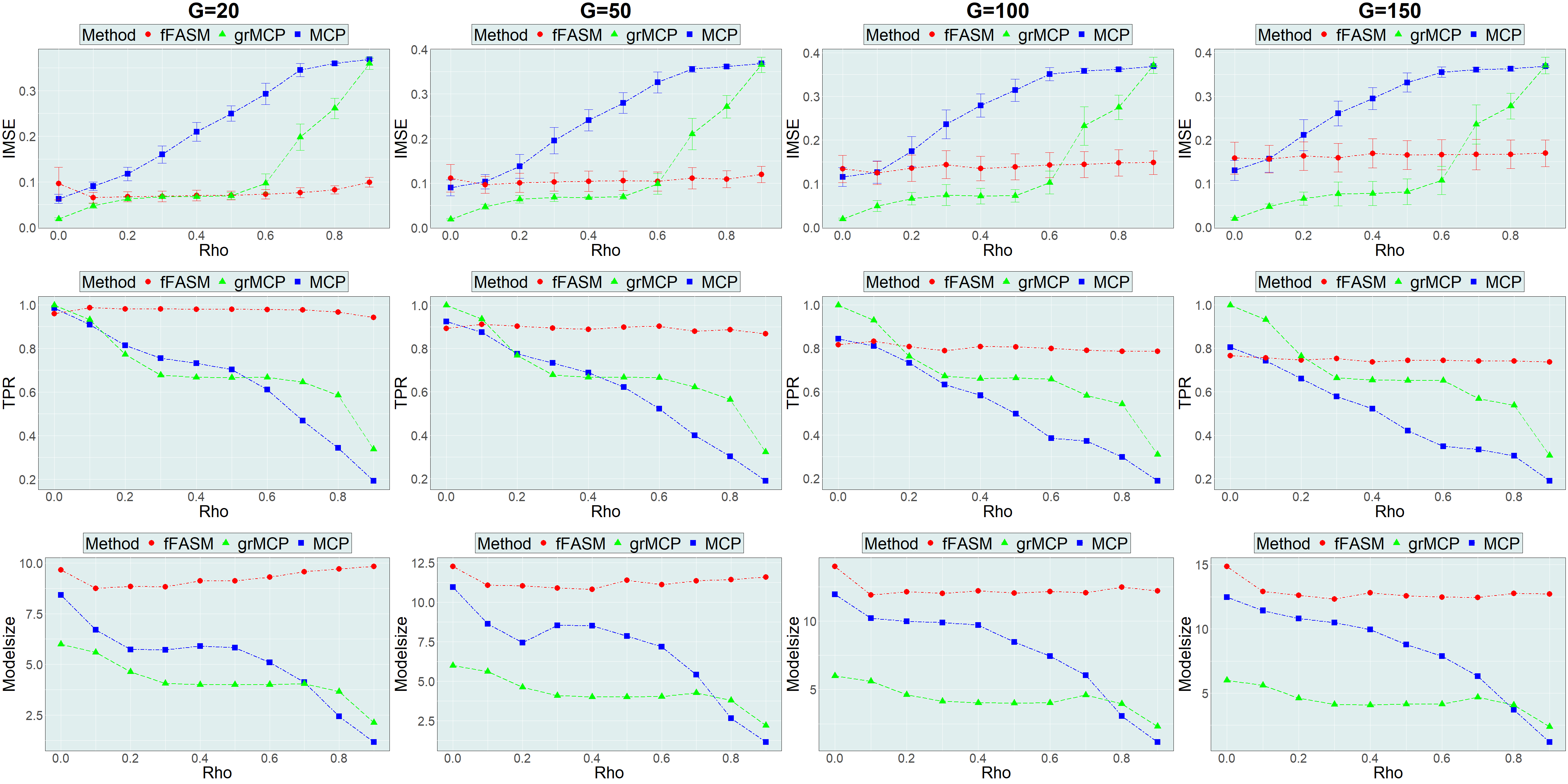}
    \caption{Estimation and selection performance of three methods with different  correlations for Scenario II. In each row, four subfigures are displayed with $G =20,50,100,150$, respectively, and in each column with a given $\rho$, the IMSE (with 0.5 standard deviations), the TPR, and model size are presented, respectively.}
    \label{fig2}
\end{figure}

Furthermore, Figure \ref{fig2} shows the estimation and selection results in Scenario II (the equal correlation case). The IMSE values of the proposed fFASM method turn out to remain at a low level when the correlation $\rho$ increases in each subfigure in the first row, indicating a robust estimation performance from fFASM, while MCP and grMCP tend to increase, demonstrating less satisfactory estimation performance. Although the two methods show slightly better results when functional covariates are of small correlation with each other, this is expected since the equal correlation indicates a small number of factors, which may be captured easily. In terms of TPR, the fFASM method is again consistently close to 1 when the correlation varies, while TPRs for the competitor methods drop dramatically. Also, when the number of functional covariates $G$ increases, the TPRs from all three methods turn out to drop without surprise. When looking at the model size, the proposed fFASM remains stable, which may slightly overestimate the model size, while the competitor models show a much less stable model size of little consistency.

\begin{figure}[htbp] 
    \centering
    \includegraphics[width=1\textwidth]{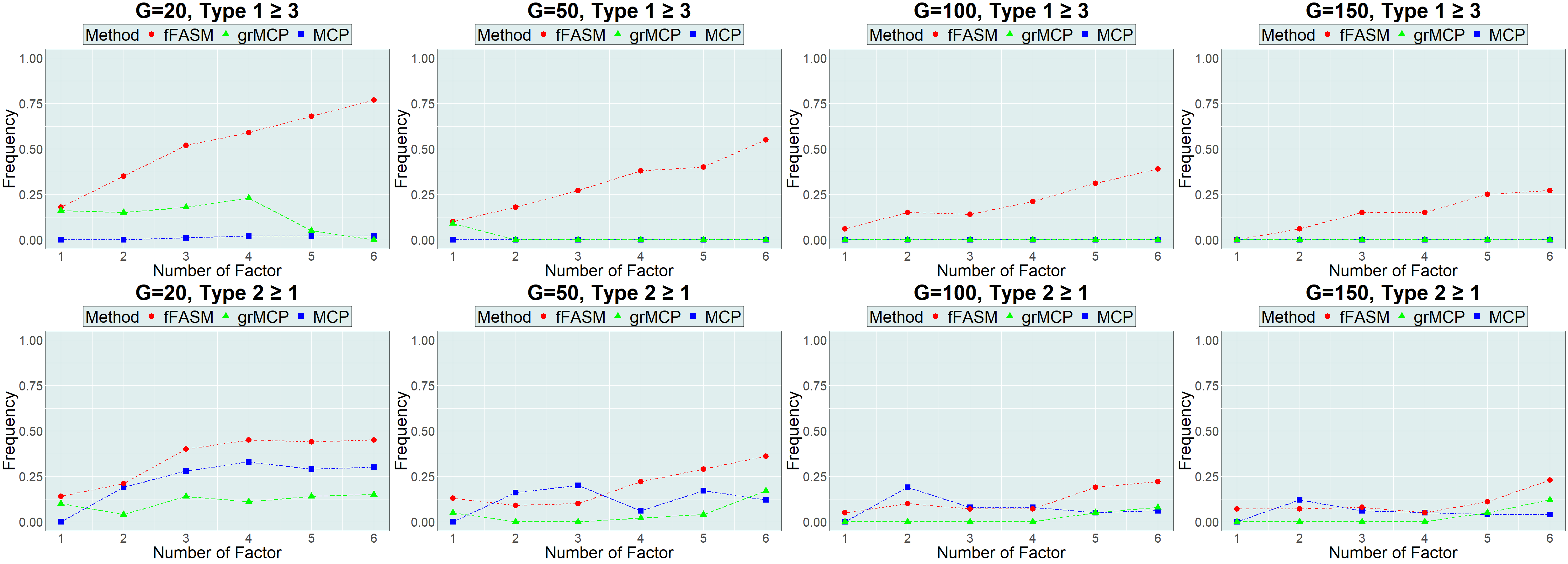}
    \caption{The selection frequencies of truly
    useful functional covariates in Scenario I with different $K$ and $G$ values. The first row displays the selection frequency that at least three out of four truly useful ones are selected for \texttt{Type1}, and the second displays the frequency that at least one out of two for \texttt{Type2}.}
    \label{fig3}
\end{figure}

\begin{figure}[htbp] 
    \centering
    \includegraphics[width=1\textwidth]{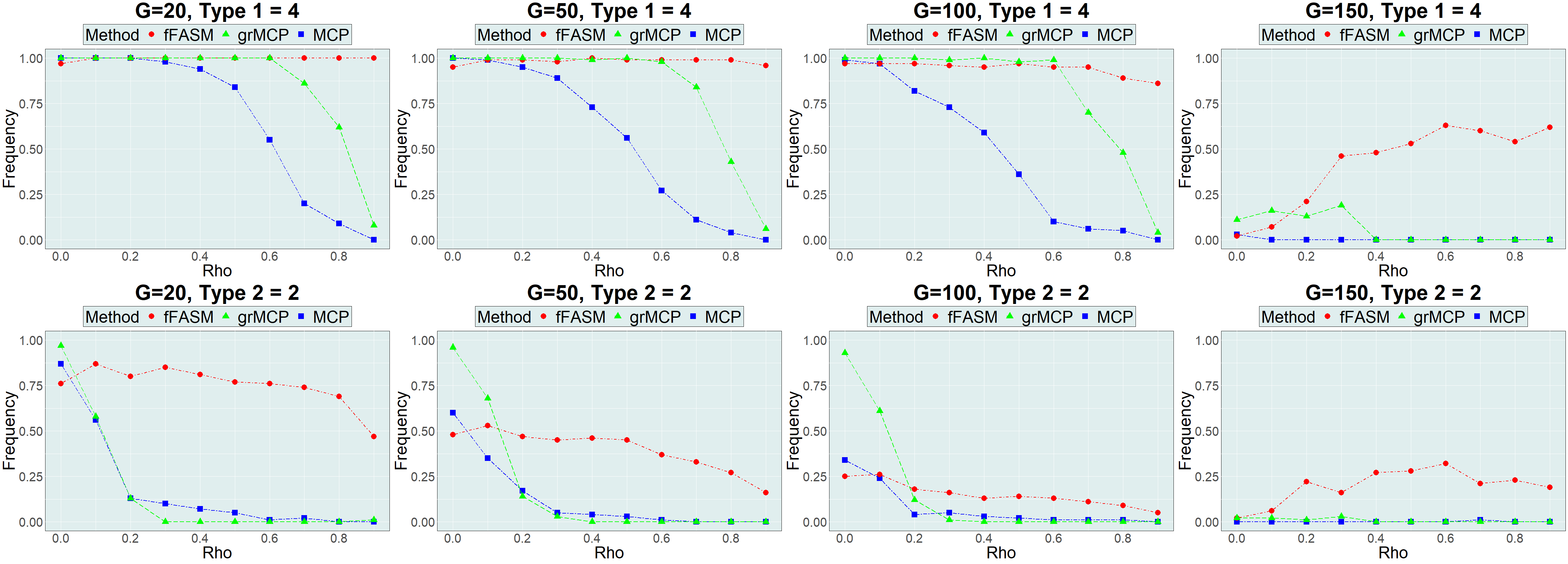}
    \caption{The selection frequencies of truly
    useful functional covariates in Scenario II with different $\rho$ and $G$ values. The first and second rows display the frequency that all truly useful ones are selected for \texttt{Type1} and \texttt{Type2}, respectively.}
    \label{fig4}
\end{figure}

Additionally, the selection frequencies of the two types of functional coefficients $\beta^{*(g)}(t)$ are examined, namely the Harmonic attenuation type (\texttt{Type1}) and the weak single signal type (\texttt{Type2}), displayed in Figure \ref{fig3} and \ref{fig4}, respectively. Note that the number of true \texttt{Type1} functional covariates is 4 and that of \texttt{Type2} is 2 in both scenarios. On one hand, Figure \ref{fig3} shows that the selection frequency of the proposed fFASM method decreases first and then increases when the number of factors $K$ increases in both two types of functional covariates, while those for MCP and grMCP tend to be 0, especially in the  {\texttt{Type2}} setting where the two competitor methods show lower starting points. Note that the selection frequencies of the proposed method for both two types are not extraordinarily high, since the functional linear model is somewhat lack of fit due to the complexity of the true model. On the other hand, in Scenario II, Figure \ref{fig4} shows a much improved selection consistency of the proposed method than grMCP and MCP for both two types, even though \texttt{Type2} may still select fewer covariates because of slight violation of norm requirement in Theorem \ref{thm1}.

\section{Real Data Application}\label{sec:Real}
\subsection{Effects of macroeconomic covariates on lifetime expectancy}
In this section, the effects of macroeconomic covariates on national expected lifetime are explored for European and American countries using the open-sourced EPS data platform \footnote{https://www.epsnet.com.cn/}. The data have been collected and documented annually over a span of 21 years, from 2000 to 2020, for 40 countries or regions. Our focus is the average lifetime expectancy as a {scalar} dependent variable against {33} macroeconomic functional covariates, such as gross domestic product, healthcare expenditure, and educational attainment. Functional covariates are correlated with each other. For example, higher levels of educational attainment often correlate with higher GDP, since a more educated workforce may usually contribute to greater economic productivity. To estimate lifetime expectancy and find out useful functional covariates, the proposed fFASM method, the MCP and grMCP methods are employed. Specifically, to recover the functional trajectories, year points from 2000 to 2022 are used as grid points. During the study, the collected data for 30 randomly selected countries (or regions) out of 40 are used as the train set to build the model, and the remaining 10 are used as the test set. 
The whole procedure is repeated 200 times, and the average out-of-sample $R^2$ and model size are reported in Table \ref{Tab.1} and the top five selected functional covariates in Table \ref{Tab.2}. 

The results from both the prediction performance in Table \ref{Tab.1} and the variable selection in Table \ref{Tab.2} may strongly support the benefits of the proposed fFASM. The average out-of-sample $R^2$ from the fFASM is the greatest among the three competitor models, indicating its competitive prediction accuracy. Further, the variables selected by fFASM are closely related to economic development and social well-being, providing insights into education and labor market conditions, which may show significant influence on the average life expectancy of a nation. Although all three methods successfully selected the key variable ``Employment'' with the highest selection frequencies, ``Adult literacy rate'' and ``total fertility rate'' are both selected with relatively large frequencies by fFASM and MCP, while grMCP fails to show high selection frequencies of them. Additionally, the fFASM method selected ``Gross national income per capita'' and ``Unemployment rate'', which are of little correlation with other variables, while MCP selected ``Labor force'' highly correlated with ``Employment'' and ``Unemployment rate''. The grMCP only focused on ``Employment'' and ignored almost all other variables with selection rates lower than 5\% .

\begin{table}[htbp] 
\centering
\begin{tabular}{cccccc}
\hline
\multicolumn{3}{c}{average out-of-sample $R^2$} & \multicolumn{3}{c}{average model size} \\ \hline
fFASM     &MCP    & grMCP  & fFASM     &MCP      & grMCP     \\ \hline
0.288   & 0.223  & 0.227  & 1.95     & 1.53      & 1.04        \\ \hline
\end{tabular}
\caption{The average out-of-sample $R^2$ and model size for EPS data.}
\label{Tab.1}
\end{table}

\begin{table}[htbp] 
\centering
\begin{tabular}{ccc}
\hline
\multirow{2}{*}{Rank} & \multicolumn{2}{c}{fFASM}              \\ \cline{2-3} 
                      & Name                      & Frequency \\ \hline
1                     & Employment (million people)  & 160         \\
2                     & Adult literacy rate (age 15 and above) (\%)                    & 56 \\ 
3                     & Total fertility rate    & 54         \\
4                     & Unemployment rate (\% of total labor force)                   & 21           \\
5                     & Gross national income per capita (current USD)                   & 17           \\
\hline
\multirow{2}{*}{Rank} & \multicolumn{2}{c}{MCP}               \\ \cline{2-3} 
                      & Name                      & Frequency \\ \hline
1                     & Employment (million people)   & 166         \\
2                     & Total fertility rate              & 57         \\
3                     & Adult literacy rate (age 15 and above, \%)     & 37         \\
4                     & Labor force (people)             & 16        \\
5                     & Unemployment rate (\% of total labor force)                    & 7         \\ \hline
\multirow{2}{*}{Rank} & \multicolumn{2}{c}{grMCP}             \\ \cline{2-3} 
                      & Name                      & Frequency \\ \hline
1                    & Employment (million people) & 187         \\
2                     & Prevalence of undernourishment (\% of total population) & 7         \\
3                     & Adult literacy rate (age 15 and above) (\%)             & 3         \\ 
4                     & Gross national income per capita (current USD)       & 2         \\
5                     & Per capita health expenditure (current USD)          & 1         \\
\hline
\end{tabular}
\caption{Top 5 variables with highest selection frequency.}
\label{Tab.2}
\end{table}

\subsection{Prediction of sales areas of commercial houses}
In this section, we focus on the prediction of annual average sales area of commercial houses in each province in China. Data are collected from the National Bureau of Statistics of China\footnote{https://data.stats.gov.cn/} on a monthly basis for 31 administrative regions of mainland China in the year 2022 and 2023. The dependent variable is the annual average sales area of commercial houses against {60} functional covariates, including cement, medium tractors, engines, aluminum materials, and other variables related to industrial output. To recover the functional trajectories, 24 grid points are predetermined on a monthly basis from 2022 to 2023. To predict the response, the proposed fFASM, the MCP and grMCP methods are employed. During the study, the collected data for 20 randomly selected provinces are used as the training set to build the model, and the remaining 11 as the test set to evaluate the prediction performance, with 200 repetitions of the whole procedure. The average out-of-sample $R^2$ and the average model sizes are reported in Table \ref{Tab.3}, and the most frequently selected functional covariates in Table \ref{Tab.4}.

From Table \ref{Tab.3}, the proposed fFASM method is found to show a better performance than MCP and grMCP in terms of a larger average out-of-sample $R^2$ and a slightly lower average model size. In Table \ref{Tab.4}, all three methods select ``Cement'' as the most frequently selected functional covariates out of 200 repetitions, which is expected as the most important raw material required for house construction. Similarly, the functional covariate ``Aluminum Materials'' is selected among top five. Specifically, the fFASM and MCP methods select the functional covariates ``Medium Tractors'' and ``Engine'', which are closely related to construction machinery used in the house building industry, while grMCP selects ``Mechanized Paper'', ``Hydropower Generation'' and ``Phosphate Rock'', which seem to have limited connection with house construction. This may explain why grMCP has a larger model size but a relatively smaller average out-of-sample performance.

\begin{table}[htbp] 
\centering
\begin{tabular}{cccccc}
\hline
\multicolumn{3}{c}{average out-of-sample $R^2$} & \multicolumn{3}{c}{average model size} \\ \hline
fFASM     &MCP    & grMCP  & fFASM     &MCP      & grMCP     \\ \hline
0.612   & 0.601  & 0.355  & 1.35     & 2.20      & 2.55        \\ \hline
\end{tabular}
\caption{The average out-of-sample $R^2$ and model size for sales area of commercial houses data.}
\label{Tab.3}
\end{table}

\begin{table}[htbp] 
\centering
\begin{tabular}{ccc}
\hline
\multirow{2}{*}{Rank} & \multicolumn{2}{c}{fFASM}              \\ \cline{2-3} 
                      & Name                      & Frequency \\ \hline
1                     & Cement  & 198         \\
2                     & Medium Tractors        & 13  \\
3                     & Engines  & 12         \\
3                     & Aluminum Materials      & 12       \\
5                     & Chemical Pesticides (Active Ingredients)    & 4         \\ \hline
\multirow{2}{*}{Rank} & \multicolumn{2}{c}{MCP}               \\ \cline{2-3} 
                      & Name                      & Frequency \\ \hline
1                     & Cement  & 199         \\
2                     & Medium Tractors     & 48         \\
3                     & Aluminum Materials  & 19         \\
4                     & Computers             & 18         \\ 
4                     & Engines           & 18  
\\ \hline
\multirow{2}{*}{Rank} & \multicolumn{2}{c}{grMCP}             \\ \cline{2-3} 
                      & Name                      & Frequency \\ \hline
1                     &  Cement & 137         \\
2                     & Aluminum Materials       & 97         \\
3                     & Mechanized Paper & 68         \\
4                     & Hydropower Generation   & 30         \\ 
5                     & Phosphate Rock         & 20         \\\hline
\end{tabular}
\caption{Top 5 variables with highest selection frequency.}
\label{Tab.4}
\end{table}

\section{Conclusion}\label{sec:Conclusion}
\label{sec:conc}

In this article, a novel functional factor augmentation structure (fFAS) is proposed to capture associations for correlated functional processes, and further a functional factor augmentation selection model (fFASM) is developed to select useful functional covariates in high dimensions with correlated functional covariates. Note only is the correlation between functional covariates addressed without assuming an explicit covariance structure, theoretical properties of the estimated functional factors are established. We primarily discuss the rationale for constructing a fFAS, how to estimate the fFAS and its estimation error, and the impact of truncating functional data on the validity and estimation of the factor model. Due to the unique characteristics of functional data, the assumed factor model and the actually estimated factor model may differ. 

Numerical investigations on both simulated and real datasets support the superior performance of the proposed fFASM method. It is found that our proposed method performs better than general functional data variable selection methods when dealing with the variable selection problem of correlated multivariate functional covariates. A practical issue may be how to determine the model size, as our method may slightly select more functional covariates in simulation studies, which may be a trade-off for modeling functional processes with correlations in high dimensions.

\bibliographystyle{agsm}
\bibliography{ref.bib}

\newpage
\bigskip
\begin{center}
{\large\bf SUPPLEMENTARY MATERIAL}
\end{center}
\section{Regularity Condition}
In this section, we introduce some regularity conditions \citep{kong2016partially}. Without loss of generality, we assume that $\{ X^{(g)}(\cdot) , g = 1 ,\ldots, G \}$ have been centred to have zero mean. 
With $W_{i}^{(g)}(t_{ij}) = X^{(g)}_i(t_{ij}) + \epsilon_{i}^{(g)}(t_{ij})$, for definiteness, we consider the local linear smoother for each set of subjects using bandwidths $\{h_{i}^{(g)}, g = 1,\ldots ,G \}$, and denote the smoothed trajectories by $ \widehat{X}_i^{(g)}(\cdot)$. 

Condition (B1) consists of regularity assumptions for functional data.
Condition (B2) is standard for local linear smoother, (B3)–(B4) concern how the functional predictors are sampled and smoothed.

\begin{itemize}
    \item[(B1)] For $g=1, \ldots, G$, for any $C>0$ there exists an $\epsilon>0$ such that
$$
\sup _{t \in \mathcal{T}}\left[\mathbb{E}\left\{\left|X^{(g)}(t)\right|^C\right\}\right]<\infty, \quad \sup _{s, t \in \mathcal{T}}\left(\mathbb{E}\left[\left\{|s-t|^{-\epsilon}\left|X^{(g)}(s)-X^{(g)}(t)\right|\right\}^C\right]\right)<\infty .
$$
For each integer $r \geq 1, \left(\tau_{k}^{(g)}\right)^{-r} E\left({a_{k}^{(g)}}\right)^{2 r}$ is bounded uniformly in $k$. 
    \item[(B2)] For $g=1, \ldots, G, X^{(g)}(\cdot)$ is twice continuously differentiable on $\mathcal{T}$ with probability 1 , and $\int \mathbb{E}\left\{X^{(g)}{}^{\prime\prime}(t)\right\}^4 d t<\infty$, where $X^{(g)}{}^{\prime\prime}(\cdot)$ denotes the second derivative of $X^{(g)}(\cdot)$.
    The following condition concerns the design on which $X_{i}^{(g)}(\cdot)$ is observed and the local linear smoother $\widehat{X}_{i}^{(g)}(\cdot)$. When a function is said to be smooth, we mean that it is continuously differentiable to an adequate order.
    \item[(B3)] For $g=1, \ldots, G,\left\{t_{i j}^{(g)}, j=1, \ldots, n_{i}^{(g)}\right\}$ are considered deterministic and ordered increasingly for $i=1, \ldots, n$. There exist densities $p_{i}^{(g)}$ uniformly smooth over $\mathcal{T}$, satisfying $\int_\mathcal{T} p_{i}^{(g)}(t) d t=1$
    and $0<c_1<\inf _i\left\{\inf _{t \in T} p_{i}^{(g)}(t)\right\}<\sup _i\left\{\sup _{t \in T} p_{i}^{(g)}(t)\right\}<c_2<\infty$ that generate $t_{i j}^{(g)}$ according to $t_{i j}^{(g)}=P_{i}^{(g)}{}^{-1}\{j /\left(n_{i}^{(g)}+1\right)\}$, where $P_{i}^{(g)}{}^{-1}$ is the inverse of $P_{i}^{(g)}(t)=\int_{-\infty}^t  p_{i}^{(g)}(s) d s$. For each $g=1, \ldots, G$, there exist a common sequence of bandwidths $h^{(g)}$ such that $0<c_1<$ $\inf _i h_{i}^{(g)} / h^{(g)}<\sup _i h_{i}^{(g)} / h^{(g)}<c_2<\infty$, where $h_{i}^{(g)}$ is the bandwidth for $\widehat{X}_{i}^{(g)}$. The kernel density function is smooth and compactly supported.

\end{itemize}
Let $\mathcal{T}=\left[a_0, b_0\right], t_{i0}^{(g)}=a_0, t_{i, n_{i}^{(g)}+1}^{(g)}=b_0$, let $\Delta_{i}^{(g)}=\sup \left\{t_{i, j+1}^{(g)}-t_{i,j}^{(g)}, j=0, \ldots, n_{i}^{(g)}\right\}$ and $n^{(g)}=n^{(g)}(n)=\inf _{i=1, \ldots, n} n_{i}^{(g)}$. The condition below is to let the smooth estimate $\widehat{X}_{i}^{(g)}$ serve as well as the true $X_{i}^{(g)}$ in the asymptotic analysis, denoting $0<\lim a_n / b_n<\infty$ by $a_n \sim b_n$.

\begin{itemize}
    \item [(B4)] For $g=1, \ldots, G, ~\sup _i \Delta_{i}^{(g)}=O\left(\left(n^{(g)}\right)^{-1}\right),~ h^{(g)} \sim \left(n^{(g)}\right)^{-1 / 5}, n^{(g)} n^{-5 / 4} \rightarrow \infty$.
\end{itemize}

\section{Proof of Theorem}
\par \textbf{Proof of Lemma 3:}
\par \textbf{Proof:} $ \Lambda_{C,K} = \operatorname{diag}\{ \lambda_1^{(C)},\ldots,\lambda_K^{(C)}\}$ consists of first $K$ eigenvalues of $C_{\bm{a}}(\omega)$. $\Lambda_{B} = \operatorname{diag}\{ \lambda_1^{(B)},\ldots,\lambda_K^{(B)}\}$ consists of first $K$ eigenvalues $BB^\top$ and note $i$-th eigenvalues of $\Lambda_u$ as $\lambda_i^{(u)}$.
\begin{align*}
    {\bm{f}}_i^\top(\omega) - \bm{f}_i^\top =& \bm{a}_i^\top(\bm{\xi}_1^{(C)}(\omega),\ldots,\bm{\xi}_K^{(C)}(\omega))\Lambda_{C,K}^{-1/2}(\omega) - (\bm{a}_i^\top - \bm{u}_i^\top)\bm{B}\Lambda_{B}^{-1} \\
    =& \bm{a}_i^\top(\bm{\xi}_1^{(C)}(\omega),\ldots,\bm{\xi}_K^{(C)}(\omega))\Lambda_{C,K}^{-1/2}(\omega) - (\bm{a}_i^\top - \bm{u}_i^\top)(\bm{\xi}_1^{(B)},\ldots,\bm{\xi}_K^{(B)})\Lambda_{B}^{-1/2} \\
    =& \bm{a}_i^\top(\bm{\xi}_1^{(B)},\ldots,\bm{\xi}_K^{(B)})(\Lambda_{C,K}^{-1/2}(\omega) - \Lambda_{B}^{-1/2}) - \bm{a}_i^\top((\mathbf{B}\mathbf{B}^\top - \lambda_k^{(B)}I)^{+}\Lambda_{\bm{u}}\bm{\xi}_k^{(B)}\omega)_{k=1,\ldots,K}\Lambda_{C,K}^{-1/2}\\
    &+\bm{u}_i^\top \mathbf{B}\Lambda_{B}^{-1} + \bm{a}_i^\top O(\omega^2).
\end{align*}
Then we give the bound separately:
For the first part, as $\lambda_i^{(C)} = \lambda_i^{(B)} + (\bm{\xi}^{(B)}_k{}^\top \Lambda_{\bm{u}} \bm{\xi}^{(B)}_k)\omega + O(\omega^2)$, we have
\begin{align*}
    \frac{1}{\sqrt{\lambda_i^{(C)}}} - \frac{1}{\sqrt{\lambda_i^{(B)}}} &= \frac{\sqrt{\lambda_i^{(B)}} - \sqrt{\lambda_i^{(C)}}}{\sqrt{\lambda_i^{(B)}\lambda_i^{(C)}}} \\
    &= \frac{{\lambda_i^{(B)}} - {\lambda_i^{(C)}}}{\sqrt{\lambda_i^{(B)}\lambda_i^{(C)}}\left( \sqrt{\lambda_i^{(B)}} + \sqrt{\lambda_i^{(C)}} \right)} \\
    &= \frac{-\left( (\bm{\xi}^{(B)}_k{}^\top \Lambda_{\bm{u}} \bm{\xi}^{(B)}_k)\omega + O(\omega^2) \right)}{\sqrt{\lambda_i^{(B)}\lambda_i^{(C)}}\left( \sqrt{\lambda_i^{(B)}} + \sqrt{\lambda_i^{(C)}} \right)}
\end{align*}

\begin{align*}
\Vert\bm{a}_i^\top(\bm{\xi}_1^{(B)},\ldots,\bm{\xi}_K^{(B)})(\Lambda_{C,K}^{-1/2}(\omega) - \Lambda_{B}^{-1/2})\Vert_2^2 &\leq 
    \Vert \vert\Lambda_{C,K}^{-1/2}(\omega) - \Lambda_{B}^{-1/2}\vert \Vert_{\max}^2
    \Vert \bm{a}_i \Vert_2^2 \\
    & \leq C_{s_1}^2 \omega^2 \Vert \bm{a}_i \Vert_2^2 + O(\omega^3)\Vert \bm{a}_i \Vert_2^2
\end{align*}
where $C_{s_1} = \frac{\lambda_{\max}(\Lambda_{\bm{u}})}{\min_{i\leq K}^{3/2}(\lambda_i^{(C)}, \lambda_i^{(B)})}$ is a constant. For the second part, 
\begin{align*}
    \vert \bm{a}_i^\top (\mathbf{B}\mathbf{B}^\top - \lambda_k^{(B)}I)^{+}\Lambda_{\bm{u}}\bm{\xi}_k^{(B)}\omega \vert^2 &\leq (\frac{\lambda_{\max}^{\bm{u}}}{\min_{i<j\leq K}(\lambda_i^{(B)} - \lambda_j^{(B)})})^2 \Vert \bm{a}_i \Vert_2^2 \omega^2 \\
    \Vert \bm{a}_i^\top((\mathbf{B}\mathbf{B}^\top - \lambda_k^{(B)}I)^{+}\Lambda_{\bm{u}}\bm{\xi}_k^{(B)}\omega)_{k=1,\ldots,K}{\Lambda^{C}}^{-1/2} \Vert_2^2 &\leq \frac{1}{\lambda^{(C)}_{\min}(\omega)}(\frac{\lambda_{\max}^{\bm{u}}}{\min_{i<j\leq K}(\lambda_i^{(B)} - \lambda_j^{(B)})})^2 \Vert \bm{a}_i \Vert_2^2 \omega^2.
\end{align*}

And the third part, with $\operatorname{Cov}(\bm{u}_i) = \omega\Lambda_u$
\begin{align*}
    \Vert \bm{u}_i^\top \mathbf{B}\Lambda_{B}^{-1} \Vert_2^2 \leq \frac{\Vert \bm{u}_i \Vert_2^2}{\lambda_{\min}^{(B)}} \leq O_P(\omega).
\end{align*}
To sum up, we obtain that
$$
 \Vert{\bm{f}}_i^\top(\omega) - \bm{f}_i^\top\Vert_2^2 = O_{P}(\omega^2\Vert \bm{a}_i \Vert_2^2 + \Vert \bm{u}_i^\top \mathbf{B}\Lambda_{B}^{-1} \Vert_2^2)
$$
For $\bm{u}_i$, with
\begin{align*}
    \bm{F} &= (\bm{A} - \bm{U})\mathbf{B}\Lambda^B{}^{-1},
\end{align*}
that we have
\begin{align*}
    \bm{U} &= \bm{A} - (\bm{A} - \bm{U})B\Lambda^B{}^{-1}\mathbf{B}^\top \\
    &= \bm{A} - (\bm{A} - \bm{U})\sum_{j = 1}^{K}\bm{\xi}_{j}{}^{(B)} \bm{\xi}_{j}^{(B)\top}.
\end{align*}
It means 
\begin{align*}
    \bm{U}\sum_{j = K+1}^{p}\bm{\xi}_{j}{}^{(B)} \bm{\xi}_{j}^{(B)\top} = \bm{A}\sum_{j = K+1}^{p}\bm{\xi}_{j}{}^{(B)} \bm{\xi}_{j}^{(B)\top}.
\end{align*}

\begin{align*}
    {\bm{u}_i}^{\top}(\omega) - \bm{u}_i^{\top} =&  \bm{a}_i^{\top}\sum_{j = K+1}^{p}\bm{\xi}_{j}^{(C)}(\omega) \bm{\xi}_{j}^{(C)}(\omega)^{\top} - \bm{a}_i^{\top}\sum_{j = K+1}^{p}\bm{\xi}_{j}^{(B)} \bm{\xi}_{j}^{(B)}{}^{\top} - \bm{u}_i^{\top}\sum_{j = 1}^{K}\bm{\xi}_{j}^{(B)} \bm{\xi}_{j}^{(B)}{}^{\top}.
\end{align*}
Where
\begin{align*}
    &  \bm{a}_i^{\top}\sum_{j = K+1}^{p}\bm{\xi}_{j}^{(C)}(\omega) \bm{\xi}_{j}^{(C)}(\omega)^{\top} - \bm{a}_i^{\top}\sum_{j = K+1}^{p}\bm{\xi}_{j}^{(B)} \bm{\xi}_{j}^{(B)}{}^{\top}\\
    =& \bm{a}_i^{\top}\left(\bm{I}_p - \sum_{j = 1}^{K}\bm{\xi}_{j}^{(C)}(\omega) \bm{\xi}_{j}^{(C)}(\omega)^{\top}\right) - \bm{a}_i^{\top} \left(\bm{I}_p - \sum_{j = 1}^{K}\bm{\xi}_{j}^{(B)} \bm{\xi}_{j}^{(B)}{}^{\top}\right) \\
    =& -\bm{a}_i^{\top} \sum_{j = 1}^{K} \bm{\xi}_{j}^{(C)}(\omega) [\bm{\xi}_{j}^{(C)}(\omega) - \bm{\xi}_{j}^{(B)}]^{\top} + \bm{a}_i^{\top} \sum_{j = 1}^{K} [\bm{\xi}_{j}^{(B)} - \bm{\xi}_{j}^{(C)}(\omega)] \bm{\xi}_{j}^{(B)}{}^{\top} \\
    =& -\bm{a}_i^{\top} \sum_{j = 1}^{K} [\bm{\xi}_{j}^{(C)}(\omega) - \bm{\xi}_{j}^{(B)}][\bm{\xi}_{j}^{(C)}(\omega) - \bm{\xi}_{j}^{(B)}]^{\top} - \bm{a}_i^{\top} \sum_{j = 1}^{K} \bm{\xi}_{j}^{(B)} [\bm{\xi}_{j}^{(C)}(\omega) - \bm{\xi}_{j}^{(B)}]^{\top}  \\
    &+ \bm{a}_i^{\top} \sum_{j = 1}^{K} [\bm{\xi}_{j}^{(B)} - \bm{\xi}_{j}^{(C)}(\omega)] \bm{\xi}_{j}^{(B)}{}^{\top} \\
    = & -\omega^2 \bm{a}_i^{\top} \sum_{j = 1}^{K} (\mathbf{B} \mathbf{B}^{\top} - \lambda_j^{(B)} I)^{+} \Lambda_u \bm{\xi}_j ^{(B)}\bm{\xi}_j ^{(B) \top} \Lambda_u (\mathbf{B} \mathbf{B}^{\top} - \lambda_j^{(B)} I)^{+} + \bm{a}_i^{\top}O(\omega^3) \\
    &+ \omega \bm{a}_i^{\top} \sum_{j = 1}^{K} \bm{\xi}_{j}^{(B)}\bm{\xi}_{j}^{(B)}{}^\top (\mathbf{B} \mathbf{B}^{\top} - \lambda_j^{(B)} I)^{+} \Lambda_u + \omega \bm{a}_i^{\top} \sum_{j = 1}^{K}  (\mathbf{B} \mathbf{B}^{\top} - \lambda_j^{(B)} I)^{+} \Lambda_u \bm{\xi}_{j}^{(B)}\bm{\xi}_{j}^{(B) \top} \\
    &+\bm{a}_i^{\top}O(\omega^2).
\end{align*}
Then we give the bound separately:
\begin{align*}
    &\|\bm{a}_i^{\top} \sum_{j = 1}^{K} (\mathbf{B} \mathbf{B}^{\top} - \lambda_j^{(B)} I)^{+} \Lambda_u \bm{\xi}_j ^{(B)}\bm{\xi}_j ^{(B) \top} \Lambda_u (\mathbf{B} \mathbf{B}^{\top} - \lambda_j^{(B)} I)^{+}\|_2^2 \\
    &\le \|\sum_{j = 1}^{K} (\mathbf{B} \mathbf{B}^{\top} - \lambda_j^{(B)} I)^{+} \Lambda_u \bm{\xi}_j ^{(B)}\bm{\xi}_j ^{(B) \top} \Lambda_u (\mathbf{B} \mathbf{B}^{\top} - \lambda_j^{(B)} I)^{+}\|_2^2 \|\bm{a}_i\|_2^2   \\
    &\le (\lambda^u_{max})^4 \|\sum_{j = 1}^{K} (\mathbf{B} \mathbf{B}^{\top} - \lambda_j^{(B)} I)^{+}  \bm{\xi}_j ^{(B)}\bm{\xi}_j ^{(B) \top} (\mathbf{B} \mathbf{B}^{\top} - \lambda_j^{(B)} I)^{+}\|_2^2 \|\bm{a}_i\|_2^2   \\
    &\le  (p-K) (\frac{\lambda_{\max}^{\bm{u}}}{\min_{i<j\leq K}(\lambda_i^{(B)} - \lambda_j^{(B)})})^4 \|\bm{a}_i\|_2^2.
\end{align*}

\begin{align*}
    \| \bm{a}_i^{\top} \sum_{j = 1}^{K} \bm{\xi}_{j}^{(B)}\bm{\xi}_{j}^{(B)}{}^\top (\mathbf{B} \mathbf{B}^{\top} - \lambda_j^{(B)} I)^{+} \Lambda_u\|_2^2 \le    (m_1+m_2-r) (\frac{\lambda_{\max}^{\bm{u}}}{\min_{i<j\leq K}(\lambda_i^{(B)} - \lambda_j^{(B)})})^2 \|\bm{a}_i\|_2^2
\end{align*}
To sum up,  remove the constant then we can obtain that
$$
 \Vert{\bm{u}}_i^\top(\omega) - \bm{u}_i^\top\Vert_2^2 = O_{P}(\omega^2 \Vert \bm{a}_i \Vert_2^2 + \Vert \bm{u}_i^\top \mathbf{B}\Lambda_{B}^{-1}\mathbf{B}^\top \Vert_2^2)
$$

\par \textbf{Proof of Lemma 4:}
\par \textbf{Proof:}

Under \textbf{(A1,A2) (B1-B4)}, using the Lemma 2(a) in \cite{kong2016partially}. 
$i = 1,\ldots,n; k_1,k_2 = 1,\ldots,s_n; j= 1,\ldots,G$
\begin{align*}
   \hat{\bm{A}}^\top\hat{\bm{A}} &= (\sum_{i=1}^n \bm{a}_i \bm{a}_i^{\top})_{G  s_n \times G  s_n} = \sum_{i=1}^n 
\left(
\begin{array}{ccc}
    \bm{a}_i^{(1)}\bm{a}_i^{(1)\top} & \ldots  & \bm{a}_i^{(1)}\bm{a}_i^{(G)\top} \\
    \ldots &\ldots  &\ldots  \\
    \bm{a}_i^{(G)}\bm{a}_i^{(1)\top} &\ldots  &  \bm{a}_i^{(G)}\bm{a}_i^{(G)\top} 
\end{array}
\right) \\
&=  \sum_{i=1}^n 
\left(
\begin{array}{ccc}
    (a_{ik_1}^{(1)}a_{ik_2}^{(1)})_{s_n \times s_n} & \ldots  & (a_{ik_1}^{(1)}a_{ik_2}^{(G)})_{s_n \times s_n} \\
    \ldots &\ldots  &\ldots  \\
    (a_{ik_1}^{(G)}a_{ik_2}^{(1)})_{s_n \times s_n} &\ldots  &  (a_{ik_1}^{(G)}a_{ik_2}^{(G)})_{s_n \times s_n}
\end{array}
\right) \\
&=   
\left(
\begin{array}{ccc}
    (\sum_{i=1}^n a_{ik_1}^{(1)}a_{ik_2}^{(1)})_{s_n \times s_n} & \ldots  & (\sum_{i=1}^n a_{ik_1}^{(1)}a_{ik_2}^{(G)})_{s_n \times s_n} \\
    \ldots &\ldots  &\ldots  \\
    (\sum_{i=1}^n a_{ik_1}^{(G)}a_{ik_2}^{(1)})_{s_n \times s_n} &\ldots  &  (\sum_{i=1}^n a_{ik_1}^{(G)}a_{ik_2}^{(G)})_{s_n \times s_n}
\end{array}
\right),
\end{align*}
then the analysis for $\{(\bm{A}^\top\bm{A})_{ij} - E(\bm{A}^\top \bm{A})_{ij}\}/n$ is is equivalent to analyzing the following .
\begin{align*}
    n^{-1}\sum_{i=1}^n [\hat{a}_{ik_1}^{(j)} \hat{a}_{ik_2}^{(j)} - 
    E( a_{ik_1}^{(j)} a_{ik_2}^{(j)})]
    &= n^{-1}\sum_{i=1}^n [(\hat{a}_{ik_1}^{(j)} \hat{a}_{ik_2}^{(j)} - a_{ik_1}^{(j)} a_{ik_2}^{(j)}) +
    (a_{ik_1}^{(j)} a_{ik_2}^{(j)} - E( a_{ik_1}^{(j)} a_{ik_2}^{(j)}))] \\
    &= O_p(k_1^{a/2 +1} n^{-1/2} + k_2^{a/2 +1} n^{-1/2} +  n^{-1/2}) \\
    &\leq O_p(s_n^{a/2 +1} n^{-1/2} ) \\
\end{align*}

\par \textbf{Proof of Theorem 1}
\par \textbf{Proof:}
We view the estimation error as perturbation
\begin{align*}
    \frac{1}{n}\bm{A}^\top\bm{A} = \mathbb{E}(\bm{a}_i \bm{a}_i^{\top}) + \frac{1}{\sqrt{n}}\Lambda_e + O(\frac{1}{n})
\end{align*}
where $\Lambda_e/\sqrt{n} + O(1/n)$ represents the estimation error, and the equation holds true with probability. $\Lambda_e$ is a bounded random matrix since $\bm{A}$ is bounded. Consider
\begin{align*}
    \widehat{\bm{f}}_i^\top - \bm{f}_i^\top(\omega) =& \widehat{\bm{a}}_i^\top(\widehat{\bm{\xi}}_1,\ldots,\widehat{\bm{\xi}}_K){\widehat{\Lambda}_K}^{-1/2} - \bm{a}_i^\top(\bm{\xi}_1^{(C)}(\omega),\ldots,\bm{\xi}_K^{(C)}(\omega))\Lambda_{C,K}^{-1/2}(\omega), \\
    \widehat{\bm{u}}_i^\top - \bm{u}_i^\top(\omega) =& \widehat{\bm{a}}_i^{\top}\sum_{j = K+1}^{p}\widehat{\bm{\xi}}_{j}\widehat{\bm{\xi}}_{j}^{\top} - \bm{a}_i^{\top}\sum_{j = K+1}^{p}\bm{\xi}_{j}^{(C)}(\omega) \bm{\xi}_{j}^{(C)}(\omega){}^\top.
\end{align*}
Follow the steps in proof of Lemma 3, we have
\begin{align*}
    \Vert \widehat{\bm{f}}_i^\top - \bm{f}_i^\top(\omega) \Vert_2^2 &\leq O_P(\frac{1}{n}\Vert \bm{a}_i \Vert_2^2) \\
    \Vert \widehat{\bm{u}}_i^\top - \bm{u}_i^\top(\omega) \Vert_2^2 &\leq O_P(\frac{1}{n}\Vert \bm{a}_i \Vert_2^2)
\end{align*}
\par \textbf{Proof of Theorem 2}
\par \textbf{Proof:}
Recall that 
\begin{equation}\label{optimization1}
\widehat{\boldsymbol{\theta}} = \left(
\begin{array}{c} 
\widehat{\mathbf{H}}_1 \\
\widehat{\mathbf{\gamma}}
\end{array}
\right) 
= \operatorname{argmin}_{\boldsymbol{\theta}}\left\{L_n(\mathbf{y}, \widehat{\mathbf{W}} \boldsymbol{\theta})+\lambda\left\|\boldsymbol{\theta}_{[p_1]}\right\|_1\right\}. 
\end{equation}
Also, Assumption \ref{asmp:estm of fm} tells us $\mathbf{V}_0$ is nonsingular and so is $\mathbf{V}=\left(\begin{array}{cc}\mathbf{I}_{p_1} & \mathbf{0}_{p_1 \times K} \\ \mathbf{0}_{K \times p_1} & \mathbf{V}_0\end{array}\right)$. 

Define $\overline{\mathbf{W}} = \left(1,\Bar{\bm{u}}_t^{\top}, \Bar{\bm{f}}_t^{\top}\right) = \widehat{\mathbf{W}} \mathbf{V} = \left(1, \widehat{\bm{u}}_t^{\top}, \widehat{\bm{f}}_t^{\top}\right) \cdot \mathbf{V}$, 
$\Bar{\boldsymbol{\theta}}=\mathbf{V}^{-1} \widehat{\boldsymbol{\theta}}$, $\widehat{\mathbf{B}}_0=\left(\mathbf{0}_K, \widehat{\mathbf{B}}^\top\right)^\top, \widehat{\boldsymbol{\theta}}^* = \left(
\begin{array}{c} 
\mathbf{H}_1^* \\
\widehat{\mathbf{B}}_0^{\top} \mathbf{H}_1^* 
\end{array}
\right)$ 
and
$\Bar{\boldsymbol{\theta}}^*=\mathbf{V}^{-1} \widehat{\boldsymbol{\theta}}^*$. We easily see that $\widehat{\mathbf{H}}=\widehat{\boldsymbol{\theta}}_{[p_1]}=\Bar{\boldsymbol{\theta}}_{[p_1]}$ and 
$$
\Bar{\boldsymbol{\theta}}=\operatorname{argmin}_{\boldsymbol{\theta}}\left\{L_n(\mathbf{y}, \overline{\mathbf{W}}\boldsymbol{\theta})+\lambda\left\|\boldsymbol{\theta}_{[p_1]}\right\|_1\right\}.
$$
Then it follows that $\operatorname{supp}(\widehat{\mathbf{H}}_1)=\operatorname{supp}\left(\Bar{\boldsymbol{\theta}}_{[p_1]}\right)$ and $\left\|\widehat{\mathbf{H}}_1-\mathbf{H}_1{}^*\right\| = \left\|\Bar{\boldsymbol{\theta}}_{[p_1]}-\Bar{\boldsymbol{\theta}}_{[p_1]}^*\right\| \leq \left\|\Bar{\boldsymbol{\theta}}-\Bar{\boldsymbol{\theta}}^*\right\|$ for any norm $\|\cdot\|$.

Consequently, Theorem \ref{thm1} is reduced to studying $\Bar{\boldsymbol{\theta}}$ and the loss function $L_n(\mathbf{y}, \overline{\mathbf{W}}\boldsymbol{\theta})$. 
The Lemma \ref{lemmaB1} below implies that all the regularity conditions (with $A=\infty$ ) in Lemma \ref{lemmaB2} are satisfied.

Let $w_{t j}$ and $\bar{w}_{t j}$ be the $(t, j)$-th element of $\mathbf{W}$ and $\overline{\mathbf{W}}$, respectively. Observe that $L_n(\mathbf{y}, \overline{\mathbf{W}} \boldsymbol{\theta})=$ $\frac{1}{n} \sum_{t=1}^n\left[-y_t \overline{\mathbf{w}}_t^\top \boldsymbol{\theta}+b\left(\overline{\mathbf{w}}_t^\top \boldsymbol{\theta}\right)\right], \nabla L_n(\mathbf{y}, \overline{\mathbf{W}} \boldsymbol{\theta})=\frac{1}{n} \sum_{t=1}^n\left[-y_t+b^{\prime}\left(\overline{\mathbf{w}}_t^T \boldsymbol{\theta}\right)\right] \overline{\mathbf{w}}_t$ and $\overline{\mathbf{W}} \Bar{\boldsymbol{\theta}}^*=\mathbf{A} \boldsymbol{H}^*+ (\beta_0^*)_{n\times 1}$. Hence $\left\|\nabla L_n\left(\mathbf{y}, \overline{\mathbf{W}} \Bar{\boldsymbol{\theta}}^*\right)\right\|_{\infty}=\varepsilon^*$ and consequently, $\left\|\nabla_S L_n\left(\mathbf{y}, \overline{\mathbf{W}} \Bar{\boldsymbol{\theta}}^*\right)\right\|_{\infty} \leq \varepsilon^*,\left\|\nabla_S L_n\left(\mathbf{y}, \overline{\mathbf{W}} \Bar{\boldsymbol{\theta}}^*\right)\right\|_2 \leq \varepsilon^* \sqrt{|S|}$ and $\left\|\nabla_S L_n\left(\mathbf{y}, \overline{\mathbf{W}} \Bar{\boldsymbol{\theta}}^*\right)\right\|_1 \leq \varepsilon^*|S|$. In addition, $\lambda>7 \varepsilon^* / \tau \geq \varepsilon^*$.

Based on these estimates, all the results follow from Theorem \ref{thm1} and some simple algebra.

Here we present the Lemma \ref{lemmaB1} used above and its proof.

\begin{lemma}(Lemma C.1 of \citet{fan2020factor})\label{lemmaB1}
Let Assumptions \ref{asmp:sm}, \ref{asmp:convexity irrepresentable} and \ref{asmp:estm of fm} hold.
Treat $L_n(\mathbf{y}, \overline{\mathbf{W}} \boldsymbol{\theta})$ as a function of $\boldsymbol{\theta}$, and the derivatives below are taken with respect to it. Firstly, by Assumption \ref{asmp:estm of fm}, we have $\|\bm{W}\|_{\max } \leq \frac{M_0}{2}$. Then
\begin{itemize}
    \item[(i)] $\left\|
    \nabla_{. S}^2 L(\mathbf{y}, \overline{\mathbf{W}} \boldsymbol{\theta})-\nabla_{. S}^2 L\left(\mathbf{y}, \overline{\mathbf{W}} \Bar{\boldsymbol{\theta}}^*\right)\right\|_{\infty} \leq M \left\| \boldsymbol{\theta} - \Bar{\boldsymbol{\theta}}^* \right\|_2, \forall \bm{\theta}$ with $\operatorname{supp}(\boldsymbol{\theta}) \subseteq S$,
    \item[(ii)] $\left\|\left(\nabla_{S S}^2 L\left(\mathbf{y}, \overline{\mathbf{W}}  \Bar{\boldsymbol{\theta}}^*\right)\right)^{-1}\right\|_{\infty} \leq \frac{1}{2 \kappa_{\infty}}$,
    \item[(iii)] $\left\|\left(\nabla_{S S}^2 L\left(\mathbf{y}, \overline{\mathbf{W}}  \Bar{\boldsymbol{\theta}}^*\right)\right)^{-1}\right\|_2 \leq \frac{1}{2 \kappa_2}$,
    \item[(iv)] $\left\|\nabla_{S_2 S}^2 L\left(\mathbf{y}, \overline{\mathbf{W}}  \Bar{\boldsymbol{\theta}}^*\right)\left(\nabla_{S S}^2 L\left(\mathbf{y}, \overline{\mathbf{W}}  \Bar{\boldsymbol{\theta}}^*\right)\right)^{-1}\right\|_{\infty} \leq 1-\tau$.
\end{itemize}
\end{lemma}
\par \textbf{Proof of Lemma 5}
(i) Based on the fact that $\mathbf{W} \boldsymbol{\theta}^*=\overline{\mathbf{W}} \overline{\boldsymbol{\theta}}^*={\bm{A}}\bm{H}^* + (\beta_0^*)_{n\times 1}$, we have $\nabla^2 L_n\left(\mathbf{y}, \mathbf{W} \boldsymbol{\theta}^*\right)=\frac{1}{n} \sum_{t=1}^n b^{\prime \prime}\left(\overline{\mathbf{w}}_t^T \overline{\boldsymbol{\theta}}^*\right) \mathbf{w}_t \mathbf{w}_t^T$ and $\nabla^2 L_n\left(\mathbf{y}, \overline{\mathbf{W}} \overline{\boldsymbol{\theta}}^*\right)=\frac{1}{n} \sum_{t=1}^n b^{\prime \prime}\left(\overline{\mathbf{w}}_t^T \overline{\boldsymbol{\theta}}^*\right) \overline{\mathbf{w}}_t \overline{\mathbf{w}}_t^T$. For any $j, k \in[p_1+K]$ and $\operatorname{supp}(\boldsymbol{\theta}) \subseteq S$,

\begin{align}\label{ieq:C1}
& \left|\nabla_{j k}^2 L_n(\mathbf{y}, \overline{\mathbf{W}} \boldsymbol{\theta})-\nabla_{j k}^2 L_n\left(\mathbf{y}, \overline{\mathbf{W}} \overline{\boldsymbol{\theta}}^*\right)\right| \leq \frac{1}{n} \sum_{t=1}^n\left|b^{\prime \prime}\left(\overline{\mathbf{w}}_t^T \boldsymbol{\theta}\right)-b^{\prime \prime}\left(\overline{\mathbf{w}}_t^T \overline{\boldsymbol{\theta}}^*\right)\right| \cdot\left|\overline{\mathbf{w}}_{t j} \overline{\mathbf{w}}_{t k}\right| \notag \\ 
& \leq \frac{1}{n} \sum_{t=1}^n M_3\left|\overline{\mathbf{w}}_t^T\left(\boldsymbol{\theta}-\overline{\boldsymbol{\theta}}^*\right)\right| \cdot\|\overline{\mathbf{W}}\|_{\max }^2
\end{align}

By the Cauchy-Schwarz inequality and $\|\overline{\mathbf{W}}\|_{\max } \leq\|\mathbf{W}\|_{\max }+\|\overline{\mathbf{W}}-\mathbf{W}\|_{\max } \leq M_{0}$, we obtain that for $i \in[n],\left|\overline{\mathbf{w}}_t^T\left(\boldsymbol{\theta}-\overline{\boldsymbol{\theta}}^*\right)\right|=\left|\overline{\mathbf{w}}_{t S}^T\left(\boldsymbol{\theta}-\overline{\boldsymbol{\theta}}^*\right)_S\right| \leq\left\|\overline{\mathbf{w}}_{t S}\right\|_2\left\|\boldsymbol{\theta}-\overline{\boldsymbol{\theta}}^*\right\|_2 \leq \sqrt{|S|} M_{0}\left\|\boldsymbol{\theta}-\overline{\boldsymbol{\theta}}^*\right\|_2$. Plugging this result back to \eqref{ieq:C1}, we get

$$
\begin{aligned}
& \left|\nabla_{j k}^2 L_n(\mathbf{y}, \overline{\mathbf{W}} \boldsymbol{\theta})-\nabla_{j k}^2 L_n\left(\mathbf{y}, \overline{\mathbf{W}} \overline{\boldsymbol{\theta}}^*\right)\right| \leq \sqrt{|S|} M_3 M_0^3\left\|\boldsymbol{\theta}-\overline{\boldsymbol{\theta}}^*\right\|_2, \quad \forall j, k \in[p+K] \\
& \left\|\nabla_{\cdot S}^2 L_n(\mathbf{y}, \overline{\mathbf{W}} \boldsymbol{\theta})-\nabla_{\cdot S}^2 L_n\left(\mathbf{y}, \overline{\mathbf{W}} \overline{\boldsymbol{\theta}}^*\right)\right\|_{\infty} \leq|S|^{3 / 2} M_3 M_0^3\left\|\boldsymbol{\theta}-\overline{\boldsymbol{\theta}}^*\right\|_2=M\left\|\boldsymbol{\theta}-\overline{\boldsymbol{\theta}}^*\right\|_2
\end{aligned}
$$

(ii) Now we come to the second claim. For any $k \in[p+K]$,

$$
\begin{aligned}
& \left\|\nabla_{k S}^2 L_n\left(\mathbf{y}, \overline{\mathbf{W}} \overline{\boldsymbol{\theta}}^*\right)-\nabla_{k S}^2 L_n\left(\mathbf{y}, \mathbf{W} \boldsymbol{\theta}^*\right)\right\|_{\infty} \leq \frac{1}{n} \sum_{t=1}^n b^{\prime \prime}\left(\mathbf{x}_t^T \boldsymbol{\beta}^*\right)\left\|\bar{w}_{t k} \overline{\mathbf{w}}_{t S}^T-w_{t k} \mathbf{w}_{t S}^T\right\|_{\infty} \\
& \leq \frac{M_2 \sqrt{|S|}}{n} \sum_{t=1}^n\left\|\bar{w}_{t k} \overline{\mathbf{w}}_{t S}^T-w_{t k} \mathbf{w}_{t S}^T\right\|_2
\end{aligned}
$$

Also, by $\|\mathbf{W}\|_{\max } \leq M_0 / 2$ and $\|\overline{\mathbf{W}}\|_{\max } \leq M_0$ we have

$$
\begin{aligned}
& \left\|\bar{w}_{t k} \overline{\mathbf{w}}_{t S}^T-w_{t k} \mathbf{w}_{t S}^T\right\|_2 \leq\left|w_{t k}\right| \cdot\left\|\left(\overline{\mathbf{w}}_{t S}-\mathbf{w}_{t S}\right)^T\right\|_2+\left|\bar{w}_{t k}-w_{t k}\right| \cdot\left\|\overline{\mathbf{w}}_{t S}^T\right\|_2 \\
& \leq\|\mathbf{W}\|_{\max }\left\|\overline{\mathbf{w}}_{t S}-\mathbf{w}_{t S}\right\|_2+\left|\bar{w}_{t k}-w_{t k}\right| \cdot \sqrt{|S|}\|\overline{\mathbf{W}}\|_{\max } \\
& \leq \frac{M_0}{2}\left\|\overline{\mathbf{w}}_{t S}-\mathbf{w}_{t S}\right\|_2+M_0 \sqrt{|S|} \cdot\left|\bar{w}_{t k}-w_{t k}\right|
\end{aligned}
$$
Define $\delta=\max _{j \in[p+K]}\left(\frac{1}{n} \sum_{t=1}^n\left|\bar{w}_{t j}-w_{t j}\right|^2\right)^{1 / 2}$. By the Jensen's inequality, $\forall J \subseteq[p+K]$,

$$
\frac{1}{n} \sum_{t=1}^n\left\|\overline{\mathbf{w}}_{t J}-\mathbf{w}_{t J}\right\|_2 \leq\left(\frac{1}{n} \sum_{t=1}^n\left\|\overline{\mathbf{w}}_{t J}-\mathbf{w}_{t J}\right\|_2^2\right)^{1 / 2} \leq\left(\frac{|J|}{n} \max _{j \in[p+K]} \sum_{t=1}^n\left|\bar{w}_{t j}-w_{t j}\right|^2\right)^{1 / 2} \leq \sqrt{|J|} \delta .
$$

As a result,

\begin{align}\label{C2}
& \left\|\nabla_{\cdot S}^2 L_n\left(\mathbf{y}, \overline{\mathbf{W}} \overline{\boldsymbol{\theta}}^*\right)-\nabla_{\cdot S}^2 L_n\left(\mathbf{y}, \mathbf{W} \boldsymbol{\theta}^*\right)\right\|_{\infty} \notag \\
& =\max _{k \in[p+K]}\left\|\nabla_{k S}^2 L_n\left(\mathbf{y}, \overline{\mathbf{W}} \overline{\boldsymbol{\theta}}^*\right)-\nabla_{k S}^2 L_n\left(\mathbf{y}, \mathbf{W} \boldsymbol{\theta}^*\right)\right\|_{\infty} \leq \frac{3}{2} M_0 M_2|S| \delta
\end{align}

Let $\alpha=\left\|\left(\nabla_{S S}^2 L_n\left(\mathbf{y}, \mathbf{W} \boldsymbol{\theta}^*\right)\right)^{-1}\left[\nabla_{S S}^2 L_n\left(\mathbf{y}, \overline{\mathbf{W}} \overline{\boldsymbol{\theta}}^*\right)-\nabla_{S S}^2 L_n\left(\mathbf{y}, \mathbf{W} \boldsymbol{\theta}^*\right)\right]\right\|_{\infty}$. Then

\begin{align}\label{C3}
\alpha & \leq\left\|\left(\nabla_{S S}^2 L_n\left(\mathbf{y}, \mathbf{W} \boldsymbol{\theta}^*\right)\right)^{-1}\right\|_{\infty}\left\|\nabla_{S S}^2 L_n\left(\mathbf{y}, \overline{\mathbf{W}}^T \overline{\boldsymbol{\theta}}^*\right)-\nabla_{S S}^2 L_n\left(\mathbf{y}, \mathbf{W} \boldsymbol{\theta}^*\right)\right\|_{\infty}\notag \\
& \leq \frac{3}{8 \kappa_{\infty}} M_0 M_2|S| \delta \leq \frac{1}{2}.
\end{align}

Lemma \ref{lemmaB2.1} yields

$$
\begin{aligned}
& \left\|\left(\nabla_{S S}^2 L_n\left(\mathbf{y}, \overline{\mathbf{W}} \overline{\boldsymbol{\theta}}^*\right)\right)^{-1}-\left(\nabla_{S S}^2 L_n\left(\mathbf{y}, \mathbf{W} \boldsymbol{\theta}^*\right)\right)^{-1}\right\|_{\infty} \leq\left\|\left(\nabla_{S S}^2 L_n\left(\mathbf{y}, \mathbf{W} \boldsymbol{\theta}^*\right)\right)^{-1}\right\|_{\infty} \frac{\alpha}{1-\alpha} \\
& \leq \frac{1}{4 \kappa_{\infty}} \cdot \frac{\alpha}{1-\frac{1}{2}} \leq \frac{3}{16 \kappa_{\infty}^2} M_0 M_2|S| \delta.
\end{aligned}
$$
We also have a cruder bound $\left\|\left(\nabla_{S S}^2 L_n\left(\mathbf{y}, \overline{\mathbf{W}} \overline{\boldsymbol{\theta}}^*\right)\right)^{-1}-\left(\nabla_{S S}^2 L_n\left(\mathbf{y}, \mathbf{W} \boldsymbol{\theta}^*\right)\right)^{-1}\right\|_{\infty} \leq \frac{1}{4 \kappa_{\infty}}$, which leads to

\begin{align}\label{C4}
    \left\|\left(\nabla_{S S}^2 L_n\left(\mathbf{y}, \overline{\mathbf{W}} \overline{\boldsymbol{\theta}}^*\right)\right)^{-1}\right\|_{\infty} \leq\left\|\left(\nabla_{S S}^2 L_n\left(\mathbf{y}, \mathbf{W} \boldsymbol{\theta}^*\right)\right)^{-1}\right\|_{\infty}+\frac{1}{4 \kappa_{\infty}} \leq \frac{1}{2 \kappa_{\infty}} 
\end{align}

(iii) The third argument follows \eqref{C4} easily. Since $\|\mathbf{A}\|_2 \leq\|\mathbf{A}\|_{\infty}$ holds for any symmetric matrix $\mathbf{A}$, we have $\left\|\left(\nabla_{S S}^2 L_n\left(\mathbf{y}, \overline{\mathbf{W}} \overline{\boldsymbol{\theta}}^*\right)\right)^{-1}-\left(\nabla_{S S}^2 L_n\left(\mathbf{y}, \mathbf{W} \boldsymbol{\theta}^*\right)\right)^{-1}\right\|_2 \leq \frac{1}{4 \kappa_{\infty}} \leq \frac{1}{4 \kappa_2}$ and thus $\left\|\left(\nabla_{S S}^2 L_n\left(\mathbf{y}, \overline{\mathbf{W}} \overline{\boldsymbol{\theta}}^*\right)\right)^{-1}\right\|_2 \leq \frac{1}{2 \kappa_2}$.
(iv) Finally we prove the last inequality. On the one hand,

$$
\begin{aligned}
& \left\|\nabla_{S_2 S}^2 L_n\left(\mathbf{y}, \overline{\mathbf{W}} \overline{\boldsymbol{\theta}}^*\right)\left(\nabla_{S S}^2 L_n\left(\mathbf{y}, \overline{\mathbf{W}} \overline{\boldsymbol{\theta}}^*\right)\right)^{-1}-\nabla_{S_2 S}^2 L_n\left(\mathbf{y}, \mathbf{W} \boldsymbol{\theta}^*\right)\left(\nabla_{S S}^2 L_n\left(\mathbf{y}, \mathbf{W} \boldsymbol{\theta}^*\right)\right)^{-1}\right\|_{\infty} \\
& \leq\left\|\nabla_{S_2 S}^2 L_n\left(\mathbf{y}, \overline{\mathbf{W}} \overline{\boldsymbol{\theta}}^*\right)-\nabla_{S_2 S}^2 L_n\left(\mathbf{y}, \mathbf{W} \boldsymbol{\theta}^*\right)\right\|_{\infty}\left\|\left(\nabla_{S S}^2 L_n\left(\mathbf{y}, \overline{\mathbf{W}} \overline{\boldsymbol{\theta}}^*\right)\right)^{-1}\right\|_{\infty} \\
& +\left\|\nabla_{S_2 S}^2 L_n\left(\mathbf{y}, \mathbf{W} \boldsymbol{\theta}^*\right)\left[\left(\nabla_{S S}^2 L_n\left(\mathbf{y}, \overline{\mathbf{W}} \overline{\boldsymbol{\theta}}^*\right)\right)^{-1}-\left(\nabla_{S S}^2 L_n\left(\mathbf{y}, \mathbf{W} \boldsymbol{\theta}^*\right)\right)^{-1}\right]\right\|_{\infty}
\end{aligned}
$$

From claim (ii) and \eqref{C2} it is easy to see that

$$
\left\|\nabla_{S_2 S}^2 L_n\left(\mathbf{y}, \overline{\mathbf{W}} \overline{\boldsymbol{\theta}}^*\right)-\nabla_{S_2 S}^2 L_n\left(\mathbf{y}, \mathbf{W} \boldsymbol{\theta}^*\right)\right\|_{\infty}\left\|\left(\nabla_{S S}^2 L_n\left(\mathbf{y}, \overline{\mathbf{W}} \overline{\boldsymbol{\theta}}^*\right)\right)^{-1}\right\|_{\infty} \leq \frac{1}{4 \kappa_{\infty}} 3 M_0 M_2|S| \delta .
$$
On the other hand, we can take $\mathbf{A}=\nabla_{S_2 S}^2 L_n\left(\mathbf{y}, \mathbf{W} \boldsymbol{\theta}^*\right), \mathbf{B}=\nabla_{S S}^2 L_n\left(\mathbf{y}, \mathbf{W} \boldsymbol{\theta}^*\right)$ and $\mathbf{C}=\nabla_{S S}^2 L_n\left(\mathbf{y}, \overline{\mathbf{W}} \overline{\boldsymbol{\theta}}^*\right)-$ $\nabla_{S S}^2 L_n\left(\mathbf{y}, \mathbf{W} \boldsymbol{\theta}^*\right)$. By Assumption \ref{asmp:convexity irrepresentable}, $\left\|\mathbf{A B}^{-1}\right\|_{\infty} \leq 1-2 \tau \leq 1$. Lemma \ref{lemmaB2.1} forces that

$$
\begin{aligned}
& \left\|\nabla_{S_2 S}^2 L_n\left(\mathbf{y}, \mathbf{W} \boldsymbol{\theta}^*\right)\left[\left(\nabla_{S S}^2 L_n\left(\mathbf{y}, \overline{\mathbf{W}} \overline{\boldsymbol{\theta}}^*\right)\right)^{-1}-\left(\nabla_{S S}^2 L_n\left(\mathbf{y}, \mathbf{W} \boldsymbol{\theta}^*\right)\right)^{-1}\right]\right\|_{\infty} \\
& =\left\|\mathbf{A}\left[(\mathbf{B}+\mathbf{C})^{-1}-\mathbf{B}^{-1}\right]\right\|_{\infty} \leq\left\|\mathbf{A} B^{-1}\right\|_{\infty} \frac{\left\|\mathbf{C B}^{-1}\right\|_{\infty}}{1-\left\|\mathbf{C B}^{-1}\right\|_{\infty}} \leq \frac{\|\mathbf{C}\|_{\infty}\left\|\mathbf{B}^{-1}\right\|_{\infty}}{1-\|\mathbf{C}\|_{\infty}\left\|\mathbf{B}^{-1}\right\|_{\infty}}
\end{aligned}
$$

We have shown above in \eqref{C3} that $\|\mathbf{C}\|_{\infty}\left\|\mathbf{B}^{-1}\right\|_{\infty} \leq \frac{3}{8 \kappa_{\infty}} M_0 M_2|S| \delta \leq 1 / 2$. As a result,

$$
\left\|\nabla_{S_2 S}^2 L_n\left(\mathbf{y}, \mathbf{W} \boldsymbol{\theta}^*\right)\left[\left(\nabla_{S S}^2 L_n\left(\mathbf{y}, \overline{\mathbf{W}} \overline{\boldsymbol{\theta}}^*\right)\right)^{-1}-\left(\nabla_{S S}^2 L_n\left(\mathbf{y}, \mathbf{W} \boldsymbol{\theta}^*\right)\right)^{-1}\right]\right\|_{\infty} \leq \frac{3}{4 \kappa_{\infty}} M_0 M_2|S| \delta.
$$

By combining these estimates, we have

$$
\begin{aligned}
& \left\|\nabla_{S_2 S}^2 L_n\left(\mathbf{y}, \overline{\mathbf{W}} \overline{\boldsymbol{\theta}}^*\right)\left(\nabla_{S S}^2 L_n\left(\mathbf{y}, \overline{\mathbf{W}} \overline{\boldsymbol{\theta}}^*\right)\right)^{-1}-\nabla_{S_2 S}^2 L_n\left(\mathbf{y}, \mathbf{W} \boldsymbol{\theta}^*\right)\left(\nabla_{S S}^2 L_n\left(\mathbf{y}, \mathbf{W} \boldsymbol{\theta}^*\right)\right)^{-1}\right\|_{\infty} \\
& \leq \frac{3}{2 \kappa_{\infty}} M_0 M_2|S| \delta \leq \tau.
\end{aligned}
$$

Therefore $\left\|\nabla_{S_2 S}^2 L_n\left(\mathbf{y}, \overline{\mathbf{W}} \overline{\boldsymbol{\theta}}^*\right)\left(\nabla_{S S}^2 L_n\left(\mathbf{y}, \overline{\mathbf{W}} \overline{\boldsymbol{\theta}}^*\right)\right)^{-1}\right\|_{\infty} \leq(1-2 \tau)+\tau=1-\tau$.

Finally, we use the following lemma to prove the theorem.
\begin{lemma}(Theorem B.1 of \citet{fan2020factor})\label{lemmaB2}
    For optimization \ref{optimization1},
    if the Lemma \ref{lemmaB1} is statisfied, then we have that
    \begin{enumerate}
        \item \textbf{Error bounds}: if
        $$
        \frac{7}{\tau}\left\|\nabla L_n\left(\boldsymbol{\theta}^*\right)\right\|_{\infty}<\lambda<\frac{\kappa_2}{4 \sqrt{|S|}} \min \left\{A, \frac{\kappa_{\infty} \tau}{3 M}\right\},
        $$
        then $\operatorname{supp}(\widehat{\boldsymbol{\theta}}) \subseteq S$ and
$$
\begin{aligned}
&\left\|\widehat{\boldsymbol{\theta}}-\boldsymbol{\theta}^*\right\|_{\infty} \leq \frac{3}{5 \kappa_{\infty}}\left(\left\|\nabla_S L_n\left(\boldsymbol{\theta}^*\right)\right\|_{\infty}+\lambda\right), \\
&\left\|\widehat{\boldsymbol{\theta}}-\boldsymbol{\theta}^*\right\|_2 \leq \frac{2}{\kappa_2}\left(\left\|\nabla_S L_n\left(\boldsymbol{\theta}^*\right)\right\|_2+\lambda \sqrt{\left|S_1\right|}\right), \\
&\left\|\widehat{\boldsymbol{\theta}}-\boldsymbol{\theta}^*\right\|_1 \leq \min \left\{\frac{3}{5 \kappa_{\infty}}\left(\left\|\nabla_S L_n\left(\boldsymbol{\theta}^*\right)\right\|_1+\lambda\left|S_1\right|\right), \frac{2 \sqrt{|S|}}{\kappa_2}\left(\left\|\nabla_S L_n\left(\boldsymbol{\theta}^*\right)\right\|_2+\lambda \sqrt{\left|S_1\right|}\right)\right\} .
\end{aligned}
$$
\item \textbf{Sign consistency }: In addition, if the following two conditions
$$
\begin{aligned}
& \min \left\{\left|{\mathbf{H}_1^*}_j\right|: {\mathbf{H}_1^*}_j \neq 0, j \in[p_1]\right\}>\frac{C}{\kappa_{\infty} \tau}\left\|\nabla L_n\left(\boldsymbol{\theta}^*\right)\right\|_{\infty} \\
& \left\|\nabla L_n\left(\boldsymbol{\theta}^*\right)\right\|_{\infty}<\frac{\kappa_2 \tau}{7 C \sqrt{|S|}} \min \left\{A, \frac{\kappa_{\infty} \tau}{3 M}\right\}
\end{aligned}
$$
hold for some $C \geq 5$, then by taking $\lambda \in\left(\frac{7}{\tau}\left\|\nabla L_n\left(\boldsymbol{\theta}^*\right)\right\|_{\infty}, \frac{1}{\tau}\left(\frac{5 C}{3}-1\right)\left\|\nabla L_n\left(\boldsymbol{\theta}^*\right)\right\|_{\infty}\right)$, the estimator achieves the sign consistency $\operatorname{sign}(\widehat{\mathbf{H}})=\operatorname{sign}\left(\mathbf{H}^*\right)$.
    \end{enumerate}
\end{lemma}
\par \textbf{Proof of Lemma 6}

We need the following two lemma to prove the Lemma \ref{lemmaB2}.

Define $B_S\left(\boldsymbol{\theta}^*, r\right)=\left\{\boldsymbol{\theta}:\left\|\boldsymbol{\theta}-\boldsymbol{\theta}^*\right\|_2 \leq r, \operatorname{supp}(\boldsymbol{\theta}) \subseteq S\right\}$ for $r>0$. We first introduce two useful lemmas.

\begin{lemma}\label{lemmaB2.1}
    Suppose $\mathbf{A} \in \mathbb{R}^{q \times r}$ and $\mathbf{B}, \mathbf{C} \in \mathbb{R}^{r \times r}$ and $\left\|\mathbf{C B}^{-1}\right\|<1$, where $\|\cdot\|$ is an induced norm. Then $\left\|\mathbf{A}\left[(\mathbf{B}+\mathbf{C})^{-1}-\mathbf{B}^{-1}\right]\right\| \leq \frac{\left\|\mathbf{A B}^{-1}\right\| \cdot\left\|\mathbf{C B}^{-1}\right\|}{1-\left\|\mathbf{C B}^{-1}\right\|}$.
\end{lemma}

\begin{lemma}\label{lemmaB2.2}
    Under (i) (ii) (iii) in Lemma \ref{lemmaB1}, we have $\left\|\left(\nabla_{S S}^2 L_n(\boldsymbol{\theta})\right)^{-1}\right\|_2<\kappa_2^{-1}$ and $\left\|\left(\nabla_{S S}^2 L_n(\boldsymbol{\theta})\right)^{-1}\right\|_{\infty}$
    $ < \kappa_{\infty}^{-1}$ over $B_S\left(\boldsymbol{\theta}^*, \min \left\{A, \frac{\kappa_{\infty}}{M}\right\}\right)$.
\end{lemma}

\begin{lemma}\label{lemmaB2.3}
    Suppose $\lambda \geq 0, \mathcal{M}$ is a Euclidean space, $\boldsymbol{\theta}_0 \in \mathcal{M}, L(\boldsymbol{\theta}) \in C^2(\mathcal{M})$ and is convex, and $R(\boldsymbol{\theta})$ is convex. In addition, there exist $\kappa, A>0$ such that $\nabla^2 L(\boldsymbol{\theta}) \succeq \kappa I$ as long as $\left\|\boldsymbol{\theta}-\boldsymbol{\theta}_0\right\|_2 \leq$ A. If $\left\|\nabla L\left(\boldsymbol{\theta}_0\right)\right\|_2+\lambda \inf _{\mathbf{h} \in \partial R\left(\boldsymbol{\theta}_0\right)}\|\mathbf{h}\|_2<\frac{1}{2} \kappa A$, then $L_\lambda(\boldsymbol{\theta})=L(\boldsymbol{\theta})+\lambda R(\boldsymbol{\theta})$ has unique minimizer $\hat{\boldsymbol{\theta}}$ and $\left\|\widehat{\boldsymbol{\theta}}-\boldsymbol{\theta}_0\right\|_2 \leq \frac{2}{\kappa}\left(\left\|\nabla L\left(\boldsymbol{\theta}_0\right)\right\|_2+\lambda \inf _{\mathbf{h} \in \partial R\left(\boldsymbol{\theta}_0\right)}\|\mathbf{h}\|_2\right)$.
\end{lemma}

\begin{lemma}\label{lemmaB2.4}
Suppose $L(\boldsymbol{\theta}) \in C^2\left(\mathbb{R}^p\right)$ and is convex. $R(\boldsymbol{\theta})$ is convex and $R(\boldsymbol{\alpha}+\boldsymbol{\beta})=R(\boldsymbol{\alpha})+$ $R(\boldsymbol{\beta})$ for $\boldsymbol{\alpha} \in \mathcal{M}$ and $\boldsymbol{\beta} \in \mathcal{M}^{\perp}$, where $\mathcal{M}$ is a linear subspace of $\mathbb{R}^p$ and $\mathcal{M}^{\perp}$ is its orthonormal complement. In addition, there exists $R^*(\boldsymbol{\theta}) \in C\left(\mathbb{R}^p\right)$ such that $|\langle\boldsymbol{\alpha}, \boldsymbol{\beta}\rangle| \leq R(\boldsymbol{\alpha}) R^*(\boldsymbol{\beta})$ for $\boldsymbol{\alpha} \in \mathcal{M}^{\perp}$ and $\boldsymbol{\beta} \in \mathbb{R}^p$. Let $L_\lambda(\boldsymbol{\theta})=L(\boldsymbol{\theta})+\lambda R(\boldsymbol{\theta})$ where $\lambda \geq 0$, and $\hat{\boldsymbol{\theta}} \in \operatorname{argmin}_{\boldsymbol{\theta} \in \mathcal{M}} L_\lambda(\boldsymbol{\theta})$.

If $R^*(\nabla L(\hat{\boldsymbol{\theta}}))<\lambda$ and $\boldsymbol{\theta}^T \nabla^2 L(\hat{\boldsymbol{\theta}}) \boldsymbol{\theta}>0$ for all $\boldsymbol{\theta} \in \mathcal{M}$, then $\hat{\boldsymbol{\theta}}$ is the unique global minimizer of $L_\lambda(\boldsymbol{\theta})$.
\end{lemma}

\par \textbf{Now we start the proof:}
First we study the restricted problem 
$$
\Bar{\boldsymbol{\theta}}=\arg\min_{\boldsymbol{\theta} \in \mathcal{M}}\left\{L_n(\boldsymbol{\theta})+\lambda R(\boldsymbol{\theta})\right\}, 
$$
where $\mathcal{M} = \{\boldsymbol{\theta}: \boldsymbol{\theta} \in \mathbb{R}^{p + K}, \boldsymbol{\theta}_{S_2} = \boldsymbol{0}\}$  is oracle parameter set.
Take $R(\boldsymbol{\theta})=\left\|\boldsymbol{\theta}_{[p]}\right\|_1$ and $R^*(\boldsymbol{\theta})=\left\|\boldsymbol{\theta}_{S_2}\right\|_{\infty}$. Let $A_1=\min \left\{A, \frac{\kappa_{\infty} \tau}{3 M}\right\}$ and hence $A_1 \leq \min \left\{A, \frac{\kappa_{\infty}}{M}\right\}$. Lemma \ref{lemmaB2.2} shows that $\left\|\left(\nabla_{S S}^2 L_n(\boldsymbol{\theta})\right)^{-1}\right\|_2<\kappa_2^{-1}$ and $\left\|\left(\nabla_{S S}^2 L_n(\boldsymbol{\theta})\right)^{-1}\right\|_{\infty}<\kappa_{\infty}^{-1}$ over $B_S\left(\boldsymbol{\theta}^*, A_1\right)$.

Since $\operatorname{supp}\left(\boldsymbol{\theta}^*\right) \subseteq S$, any $\mathbf{h} \in \partial R\left(\boldsymbol{\theta}^*\right)$ satisfies $\|\mathbf{h}\|_2 \leq \sqrt{\left|S_1\right|}$. Therefore
$$
\left\|\nabla_S L_n\left(\boldsymbol{\theta}^*\right)\right\|_2+\lambda\|\mathbf{h}\|_2 \leq \frac{1}{2} \kappa_2 A_1 \leq \frac{1}{2} \kappa_2 A .
$$
Then Lemma \ref{lemmaB2.3} implies that $\left\|\Bar{\boldsymbol{\theta}}-\boldsymbol{\theta}^*\right\|_2 \leq \frac{2}{\kappa_2}\left(\left\|\nabla_S L\left(\boldsymbol{\theta}^*\right)\right\|_2+\lambda \sqrt{\left|S_1\right|}\right) \leq A_1$.

Second, we study the $L_{\infty}$ bound. On the one hand, the optimality condition yields $\nabla_S L_n(\Bar{\boldsymbol{\theta}}) \in -\lambda \partial\left\|\Bar{\boldsymbol{\theta}}_{S}\right\|_{\infty}$ and hence $\left\|\nabla_S L_n(\Bar{\boldsymbol{\theta}})\right\|_{\infty} \leq \lambda$. On the other hand, by letting $\boldsymbol{\theta}_t=(1-t) \boldsymbol{\theta}^*+t \Bar{\boldsymbol{\theta}}(0 \leq t \leq 1)$ we have
$$
\begin{aligned}
& \nabla_S L_n(\Bar{\boldsymbol{\theta}})-\nabla_S L_n\left(\boldsymbol{\theta}^*\right)=\int_0^1 \nabla_{S S}^2 L_n\left(\boldsymbol{\theta}_t\right)\left(\Bar{\boldsymbol{\theta}}-\boldsymbol{\theta}^*\right)_{S} \mathrm{d} t \\
& =\nabla_{S S}^2 L_n\left(\boldsymbol{\theta}^*\right)\left(\Bar{\boldsymbol{\theta}}-\boldsymbol{\theta}^*\right)_S+\int_0^1\left[\nabla_{S S}^2 L_n\left(\Bar{\boldsymbol{\theta}}_t\right)_{S}-\nabla_{S S}^2 L_n\left(\boldsymbol{\theta}^*\right)\right]\left(\Bar{\boldsymbol{\theta}}-\boldsymbol{\theta}^*\right)_{S} \mathrm{d} t .
\end{aligned}
$$
Hence
$$
\begin{aligned}
& \left\|\left(\Bar{\boldsymbol{\theta}}-\boldsymbol{\theta}^*\right)_S-\left(\nabla_{S S}^2 L_n\left(\boldsymbol{\theta}^*\right)\right)^{-1}\left[\nabla_S L_n(\Bar{\boldsymbol{\theta}})-\nabla_S L_n\left(\boldsymbol{\theta}^*\right)\right]\right\|_{\infty} \\
& \leq \int_0^1\left\|\left(\nabla_{S S}^2 L_n\left(\boldsymbol{\theta}^*\right)\right)^{-1}\left[\nabla_{S S}^2 L_n\left(\Bar{\boldsymbol{\theta}}_t\right)-\nabla_{S S}^2 L_n\left(\boldsymbol{\theta}^*\right)\right]\left(\Bar{\boldsymbol{\theta}}-\boldsymbol{\theta}^*\right)_S\right\|_{\infty} \mathrm{d} t \\
& \leq\left\|\left(\nabla_{S S}^2 L_n\left(\boldsymbol{\theta}^*\right)\right)^{-1}\right\|_{\infty} \sup _{t \in[0,1]}\left\|\nabla_{S S}^2 L_n\left(\Bar{\boldsymbol{\theta}}_t\right)-\nabla_{S S}^2 L_n\left(\boldsymbol{\theta}^*\right)\right\|_{\infty}\left\|\Bar{\boldsymbol{\theta}}-\boldsymbol{\theta}^*\right\|_{\infty}
\end{aligned}
$$
By (i) (ii) (iii) in Lemma \ref{lemmaB1}, we obtain that
$$
\left\|\left(\Bar{\boldsymbol{\theta}}-\boldsymbol{\theta}^*\right)_S-\left(\nabla_{S S}^2 L_n\left(\boldsymbol{\theta}^*\right)\right)^{-1}\left[\nabla_S L_n(\Bar{\boldsymbol{\theta}})-\nabla_S L_n\left(\boldsymbol{\theta}^*\right)\right]\right\|_{\infty} \leq \frac{M}{2 \kappa_{\infty}}\left\|\Bar{\boldsymbol{\theta}}-\boldsymbol{\theta}^*\right\|_2\left\|\Bar{\boldsymbol{\theta}}-\boldsymbol{\theta}^*\right\|_{\infty} .
$$
By $\Bar{\boldsymbol{\theta}} \in B_S\left(\boldsymbol{\theta}^*, A_1\right)$ we have
$$
\begin{aligned}
\left\|\Bar{\boldsymbol{\theta}}-\boldsymbol{\theta}^*\right\|_{\infty} &\leq\left\|\left(\nabla_{S S}^2 L_n\left(\boldsymbol{\theta}^*\right)\right)^{-1}\right\|_{\infty}\left\|\nabla_S L_n(\Bar{\boldsymbol{\theta}})-\nabla_S L_n\left(\boldsymbol{\theta}^*\right)\right\|_{\infty}+\frac{M}{2 \kappa_{\infty}}\left\|\Bar{\boldsymbol{\theta}}-\boldsymbol{\theta}^*\right\|_2\left\|\Bar{\boldsymbol{\theta}}-\boldsymbol{\theta}^*\right\|_{\infty} \\
& \leq \frac{1}{2 \kappa_{\infty}}\left(\lambda+\left\|\nabla_S L_n\left(\boldsymbol{\theta}^*\right)\right\|_{\infty}\right)+\frac{1}{6}\left\|\overline{\boldsymbol{\theta}}-\boldsymbol{\theta}^*\right\|_{\infty} .
\end{aligned}
$$
Therefore,
\begin{equation}\label{eq:bound_inf}
\| \Bar{\boldsymbol{\theta}}-\boldsymbol{\theta}^* \|_{\infty} \leq \frac{3}{5 \kappa_{\infty}}\left(\left\|\nabla_S L_n\left(\boldsymbol{\theta}^*\right)\right\|_{\infty}+\lambda\right) .
\end{equation}

Third we study the $L_1$ bound. The bound on $\left\|\Bar{\boldsymbol{\theta}}-\boldsymbol{\theta}^*\right\|_1$ can be obtained in a similar way. Using the fact that $\|\cdot\|_1=\|\cdot\|_{\infty}$ for symmetric matrices,
$$
\begin{aligned}
 \left\|\Bar{\boldsymbol{\theta}}-\boldsymbol{\theta}^*\right\|_1 &\leq\left\|\left(\nabla_{S S}^2 L_n\left(\boldsymbol{\theta}^*\right)\right)^{-1}\right\|_1\left\|\nabla_S L_n(\Bar{\boldsymbol{\theta}})-\nabla_S L_n\left(\boldsymbol{\theta}^*\right)\right\|_1+\frac{M}{2 \kappa_{\infty}}\left\|\Bar{\boldsymbol{\theta}}-\boldsymbol{\theta}^*\right\|_2\left\|\Bar{\boldsymbol{\theta}}-\boldsymbol{\theta}^*\right\|_1 \\
& \leq \frac{1}{2 \kappa_{\infty}}\left(\lambda\left|S_1\right|+\left\|\nabla_S L_n\left(\boldsymbol{\theta}^*\right)\right\|_1\right)+\frac{1}{6}\left\|\Bar{\boldsymbol{\theta}}-\boldsymbol{\theta}^*\right\|_1 .
\end{aligned}
$$
Hence $\left\|\Bar{\boldsymbol{\theta}}-\boldsymbol{\theta}^*\right\|_1 \leq \frac{3}{5 \kappa_{\infty}}\left(\left\|\nabla_S L_n\left(\boldsymbol{\theta}^*\right)\right\|_1+\lambda\left|S_1\right|\right)$. Since $\operatorname{supp}(\Bar{\boldsymbol{\theta}}) \subseteq S$, we also have
$$
\left\|\Bar{\boldsymbol{\theta}}-\boldsymbol{\theta}^*\right\|_1 \leq \sqrt{|S|}\left\|\Bar{\boldsymbol{\theta}}-\boldsymbol{\theta}^*\right\|_2 \leq \frac{2 \sqrt{|S|}}{\kappa_2}\left(\left\|\nabla_S L\left(\boldsymbol{\theta}^*\right)\right\|_2+\lambda \sqrt{\left|S_1\right|}\right) .
$$

This gives another $L_1$ bound.
By Lemma \ref{lemmaB2.4}, to derive $\hat{\boldsymbol{\theta}}=\Bar{\boldsymbol{\theta}}$ it remains to show that $\left\|\nabla_{S_2} L_n(\Bar{\boldsymbol{\theta}})\right\|_{\infty}<\lambda$. Using the Taylor expansion we have
\begin{equation}\label{eq:lemma6}
\begin{aligned}
& \nabla_{S_2} L_n(\Bar{\boldsymbol{\theta}})-\nabla_{S_2} L_n\left(\boldsymbol{\theta}^*\right)=\int_0^1 \nabla_{S_2 S}^2 L_n\left(\boldsymbol{\theta}_t\right)\left(\Bar{\boldsymbol{\theta}}-\boldsymbol{\theta}^*\right)_S \mathrm{d} t \\
& =\nabla_{S_2 S}^2 L_n\left(\boldsymbol{\theta}^*\right)\left(\Bar{\boldsymbol{\theta}}-\boldsymbol{\theta}^*\right)+\int_0^1\left[\nabla_{S_2 S}^2 L_n\left(\boldsymbol{\theta}_t\right)-\nabla_{S_2 S}^2 L_n\left(\boldsymbol{\theta}^*\right)\right]\left(\Bar{\boldsymbol{\theta}}-\boldsymbol{\theta}^*\right)_S \mathrm{d} t .
\end{aligned}
\end{equation}
On the one hand, the first term in \eqref{eq:lemma6} follows,
$$
\begin{aligned}
 \left\|\nabla_{S_2 S}^2 L_n\left(\boldsymbol{\theta}^*\right)\left(\Bar{\boldsymbol{\theta}}-\boldsymbol{\theta}^*\right)\right\|_{\infty}&=\left\|\left[\nabla_{S_2 S}^2 L_n\left(\boldsymbol{\theta}^*\right)\left(\nabla_{S S}^2 L_n\left(\boldsymbol{\theta}^*\right)\right)^{-1}\right]\left[\nabla_{S S}^2 L_n\left(\boldsymbol{\theta}^*\right)\left(\Bar{\boldsymbol{\theta}}-\boldsymbol{\theta}^*\right)\right]\right\|_{\infty} \\
& \leq(1-\tau)\left\|\nabla_{S S}^2 L_n\left(\boldsymbol{\theta}^*\right)\left(\Bar{\boldsymbol{\theta}}-\boldsymbol{\theta}^*\right)\right\|_{\infty} .
\end{aligned}
$$

By the Taylor expansion, triangle's inequality, (i) in Lemma \ref{lemmaB1} and the fact that $\Bar{\boldsymbol{\theta}} \in B_S\left(\boldsymbol{\theta}^*, A_1\right)$,
$$
\begin{aligned}
\left\|\nabla_{S S}^2 L_n\left(\boldsymbol{\theta}^*\right)\left(\Bar{\boldsymbol{\theta}}-\boldsymbol{\theta}^*\right)\right\|_{\infty} 
&\leq\left\|\nabla_S L_n(\Bar{\boldsymbol{\theta}})-\nabla_S L_n\left(\boldsymbol{\theta}^*\right)\right\|_{\infty}\\
&+\int_0^1\left\|\left[\nabla_{S S}^2 L_n\left(\Bar{\boldsymbol{\theta}}_t\right)-\nabla_{S S}^2 L_n\left(\boldsymbol{\theta}^*\right)\right]\left(\Bar{\boldsymbol{\theta}}-\boldsymbol{\theta}^*\right)\right\|_{\infty} \mathrm{d} t \\
& \leq\left\|\nabla_S L_n(\Bar{\boldsymbol{\theta}})\right\|_{\infty}+\left\|\nabla_S L_n\left(\boldsymbol{\theta}^*\right)\right\|_{\infty}+M\left\|\Bar{\boldsymbol{\theta}}-\boldsymbol{\theta}^*\right\|_2\left\|\Bar{\boldsymbol{\theta}}-\boldsymbol{\theta}^*\right\|_{\infty} \\
& \leq \lambda+\left\|\nabla_S L_n\left(\boldsymbol{\theta}^*\right)\right\|_{\infty}+\frac{\kappa_{\infty} \tau}{3}\left\|\Bar{\boldsymbol{\theta}}-\boldsymbol{\theta}^*\right\|_{\infty} .
\end{aligned}
$$

On the other hand, we bound the second term in (B.4). Note that $\boldsymbol{\theta}_t \in B_S\left(\boldsymbol{\theta}^*, A_1\right)$ for all $t \in[0,1]$. (i) in Lemma \ref{lemmaB1} yields
$$
\begin{aligned}
& \left\|\int_0^1\left[\nabla_{S_2 S}^2 L_n\left(\boldsymbol{\theta}_t\right)-\nabla_{S_2 S}^2 L_n\left(\boldsymbol{\theta}^*\right)\right]\left(\Bar{\boldsymbol{\theta}}-\boldsymbol{\theta}^*\right) \mathrm{d} t\right\|_{\infty} \\
& \leq \sup _{t \in[0,1]}\left\|\nabla_{S_2 S}^2 L_n\left(\boldsymbol{\theta}_t\right)-\nabla_{S_2 S}^2 L_n\left(\boldsymbol{\theta}^*\right)\right\|_{\infty}\left\|\Bar{\boldsymbol{\theta}}-\boldsymbol{\theta}^*\right\|_{\infty} \leq \frac{\kappa_{\infty} \tau}{3}\left\|\Bar{\boldsymbol{\theta}}-\boldsymbol{\theta}^*\right\|_{\infty} .
\end{aligned}
$$

As a result,
$$
\begin{aligned}
\left\|\nabla_{S_2} L_n(\overline{\boldsymbol{\theta}})\right\|_{\infty} &\leq\left\|\nabla_{S_2} L_n\left(\boldsymbol{\theta}^*\right)\right\|_{\infty}+(1-\tau)\left(\lambda+\left\|\nabla_S L_n\left(\boldsymbol{\theta}^*\right)\right\|_{\infty}+\frac{\kappa_{\infty} \tau}{3}\left\|\Bar{\boldsymbol{\theta}}-\boldsymbol{\theta}^*\right\|_{\infty}\right)\\
&+\frac{\kappa_{\infty} \tau}{3}\left\|\Bar{\boldsymbol{\theta}}-\boldsymbol{\theta}^*\right\|_{\infty} \\
& \leq \lambda-\tau\left(\lambda-\frac{2 \kappa_{\infty}}{3}\left\|\Bar{\boldsymbol{\theta}}-\boldsymbol{\theta}^*\right\|_{\infty}-\frac{2}{\tau}\left\|\nabla L_n\left(\boldsymbol{\theta}^*\right)\right\|_{\infty}\right) .
\end{aligned}
$$
Recall that the $L_{\infty}$ bound in \eqref{eq:bound_inf}. By plugging in this estimate, and using the assumptions $0<\tau<1$ and $\lambda>\frac{20}{3 \tau}\left\|\nabla L_n\left(\boldsymbol{\theta}^*\right)\right\|_{\infty}$, we derive that
$$
\begin{aligned}
\left\|\nabla_{S_2} L_n(\Bar{\boldsymbol{\theta}})\right\|_{\infty} &\leq \lambda-\tau\left(\lambda-\frac{2}{5}\left(\left\|\nabla_S L_n\left(\boldsymbol{\theta}^*\right)\right\|_{\infty}+\lambda\right)-\frac{2}{\tau}\left\|\nabla L_n\left(\boldsymbol{\theta}^*\right)\right\|_{\infty}\right) \\
& \leq \lambda-\tau\left(\frac{3}{5} \lambda-\frac{4}{\tau}\left\|\nabla L_n\left(\boldsymbol{\theta}^*\right)\right\|_{\infty}\right)<\lambda .
\end{aligned}
$$

This implies $\widehat{\boldsymbol{\theta}}=\Bar{\boldsymbol{\theta}}$ and translates all the bounds for $\Bar{\boldsymbol{\theta}}$ to the ones for $\widehat{\boldsymbol{\theta}}$. The proposition on sign consistency follows from elementary computation, thus we omit its proof.

\par \textbf{Proof of Theorem 3}
\par \textbf{Proof:}
Write the loss function \ref{loss:g} as $L_n\left(\bm{y},\widehat{\bm{W}}\bm{\theta}\right) = \frac{1}{n}\sum_{i=1}^n L\left(y_i,\widehat{\bm{w}}_i^\top \bm{\theta}\right)$. We note $L\left(y_i,\widehat{\bm{w}}_i^\top \bm{\theta}\right)|_{\bm{A}}$ to represent that $\widehat{\bm{f}}_i$ and $\widehat{\bm{u}}_i$ of the loss function are calculated by $\bm{A}$, so as $L\left(y_i,\widehat{\bm{w}}_i^\top \bm{\theta}\right)|_{\widehat{\bm{A}}}$. Note $\zeta_i = L\left(y_i,\widehat{\bm{w}}_i^\top \bm{\theta}\right)|_{\widehat{\bm{A}}} - L\left(y_i,\widehat{\bm{w}}_i^\top \bm{\theta}\right)|_{{\bm{A}}} $. For a fixed $n$, $\zeta_i, i=1,\ldots,n$ have the same distribution.
\begin{align*}
    \text{Var}(\frac{1}{n}\sum_{i=1}^n \zeta_i)\leq \mathbb{E}(\zeta_i^2).
\end{align*}
With $\zeta_i \rightarrow 0$ in probability, we have $\mathbb{E}(\zeta_i^2) \rightarrow 0$.

\end{document}